\documentclass[letterpaper, 10 pt, conference]{ieeeconf}  

\IEEEoverridecommandlockouts                              

\usepackage{array}
\usepackage{graphicx}
\usepackage{subcaption}
\usepackage[english]{babel}
\usepackage{xcolor}

\title{\LARGE \bf
A transfer-learning approach for lesion detection \\
in endoscopic images from the urinary tract
}

\author{Jorge F. Lazo$^{1,2}$, 
Sara Moccia$^{3,4}$, 
Aldo Marzullo$^{5}$,  
Michele Catellani$^{6}$,\\ 
Ottavio de Cobelli$^{6}$,
Benoit Rosa$^{2}$, 
Michel de Mathelin$^{2}$, 
Elena De Momi$^{1}$ \\
\textit{$^{1}$ Department of Electronics, Information and Bioengineering}, Politecnico di Milano, Milan, Italy \\
\textit{$^2$ ICube, UMR 7357}, CNRS-Université de Strasbourg, Strasbourg, France \\
\textit{$^3$ The BioRobotics Institute}, Scuola Superiore Sant’Anna, Pisa, Italy\\
\textit{$^4$ Department of Excellence in Robotics and AI}, Scuola Superiore Sant’Anna, Pisa, Italy\\
\textit{$^5$ Department of Mathematics and Computer Science}, University of Calabria, Rende (CS), Italy. \\
\textit{$^6$ Istituto Europeo di Oncologia (IRCCS)}, Milan, Italy\\

\thanks{
This work was supported by the ATLAS project. This project has received funding from the European Union’s Horizon 2020 research and innovation programme under the Marie Skłodowska-Curie grant agreement No 813782.
}
}

\begin{document}

\maketitle
\thispagestyle{empty}
\pagestyle{empty}

\begin{abstract}
\indent Ureteroscopy and cystoscopy are the gold standard methods to identify and treat tumors along the urinary tract. 
It has been reported that during a normal procedure a rate of 10-20 $\%$ of the lesions could be missed. 
In this work we study the implementation of 3 different Convolutional Neural Networks (CNNs), using a 2-steps training strategy, to classify images from the urinary tract with and without lesions. 
A total of 6,101 images from ureteroscopy and cystoscopy procedures were collected. 
The CNNs were trained and tested using transfer learning in a two-steps fashion on 3 datasets. 
The datasets used were: 1)~only ureteroscopy images, 2)~only cystoscopy images and 3)~the combination of both of them. 
For cystoscopy data, VGG performed better obtaining an Area Under the ROC Curve (AUC) value of $0.846$. 
In the cases of ureteroscopy and the combination of both datasets, ResNet50 achieved the better results with AUC values of $0.987$ and $0.940$. 
The use of a training dataset which comprehends both domains results in general better performances, but performing a second stage of transfer learning achieves comparable ones. 
There is no single model which performs better in all scenarios, but ResNet50 is the network that achieves the better performances in most of them. 
The obtained results open the opportunity for further investigation with a view for improving lesion detection in endoscopic images of the urinary system.

\indent \textit{Clinical relevance} — A computer-assisted method based on CNNs could lead to a better early and adequate detection of tumor-like lesions in the urinary tract. This could support surgeons to perform better follow-ups and reduce the high recurrence rates present on this disease. 

\end{abstract}

\section{INTRODUCTION}
Urinary tract cancer is common and comprises different types of lesions ranging from small benign tumors to aggressive neoplasms with high mortality.  
In 2020, this disease had 159,120 patients affected in the United States~\cite{siegel2020cancer}. 
Urothelial carcinoma can occur in kidney pelvis, ureters, urinary bladder, and urethra. 
The predominant urinary tract malignancy is Bladder Cancer (BC), which represents more than 90$\%$ of the cases~\cite{kausch2010photodynamic}. 

Ureteroscopy (URS) and Cystoscopy (CYS) dare the main techniques for upper tract urothelial cancer (UTUC) and BC diagnosis and treatment~\cite{cosentino2013upper}. 
The procedures are carried out with the visual guidance of an endoscopic camera~\cite{wason2020ureteroscopy} in most cases using White Light Imaging (WLI) or Narrow Band Imaging (NBI)~\cite{zheng2012narrow} . 

The clinical challenge for BC is to reduce the high recurrence rate with values of $75\%$~\cite{sylvester2006predicting} and the miss rate, reported to be between 10-20$\%$~\cite{chou2017comparative}.
Computer-aided diagnosis (CAD) systems for endoscopic image analysis have recently shown to be able to support clinicians diagnosis capabilities during endoscopic procedures~\cite{poon2020ai}. 
Deep Learning (DL) methods and specifically Convolutional Neural Networks (CNNs) have become in the last years the standard for image analysis in endoscopic image data~\cite{patrini2020transfer}.

\newcommand{\widthfig}{1.9cm}
\newcommand{\heighfig}{2.1cm}
\begin{figure}[tbp]
    \begin{center}
    \begin{subfigure}[b]{0.23\textwidth}
    \centering
        \includegraphics[width=\widthfig, height=\heighfig]{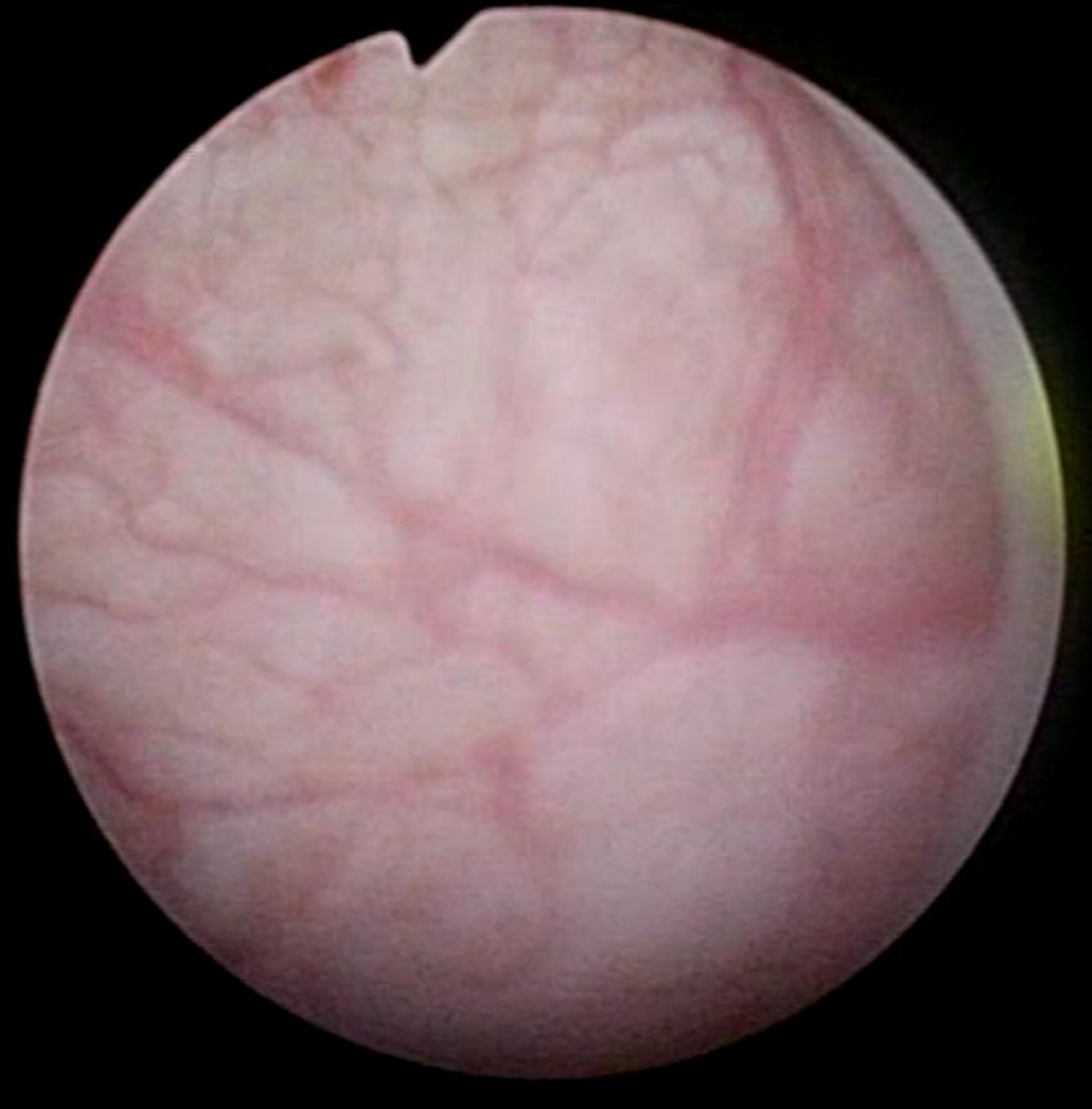} 
        \includegraphics[width=\widthfig, height=\heighfig]{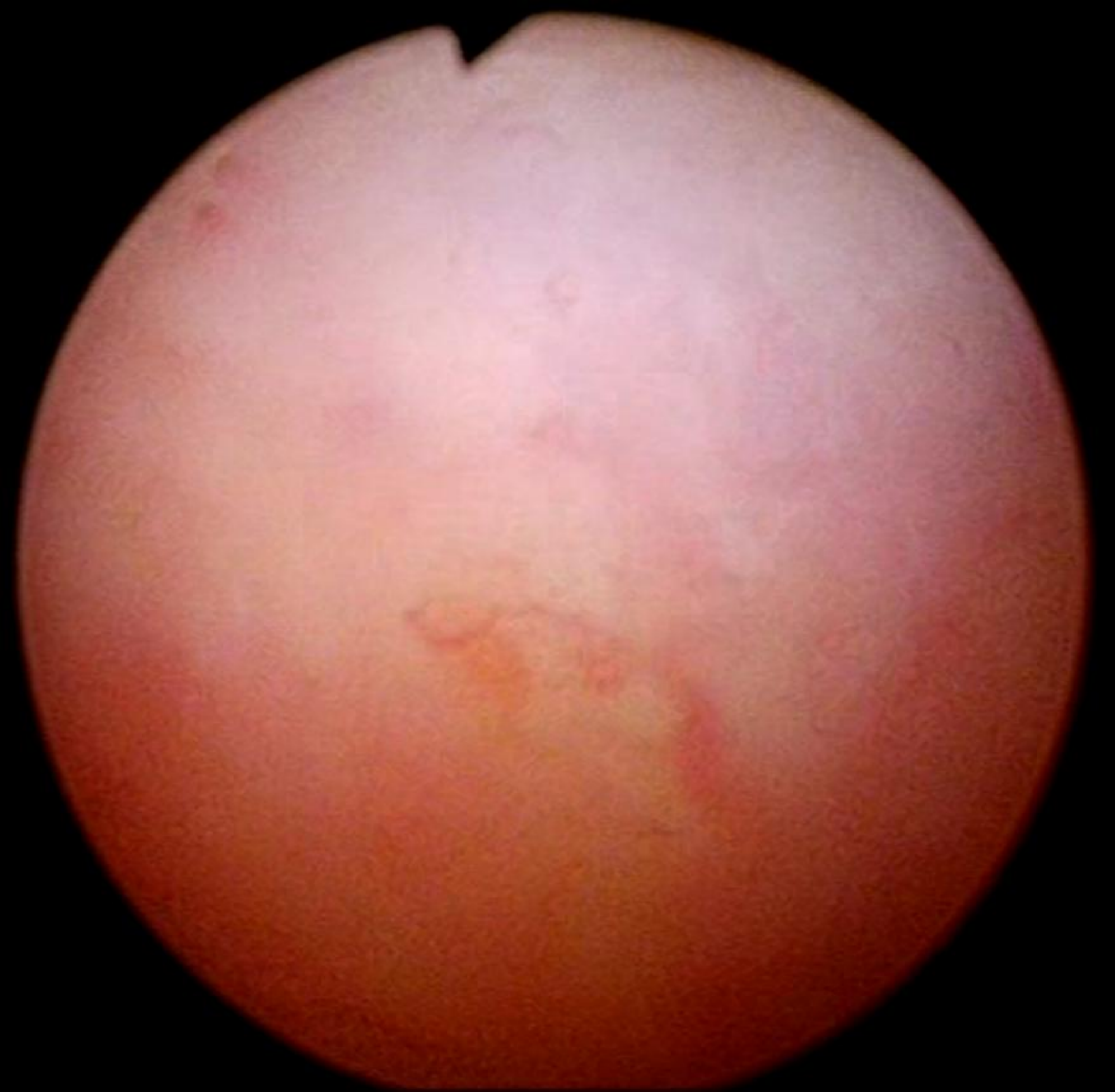} 
          
        \includegraphics[width=\widthfig, height=\heighfig]{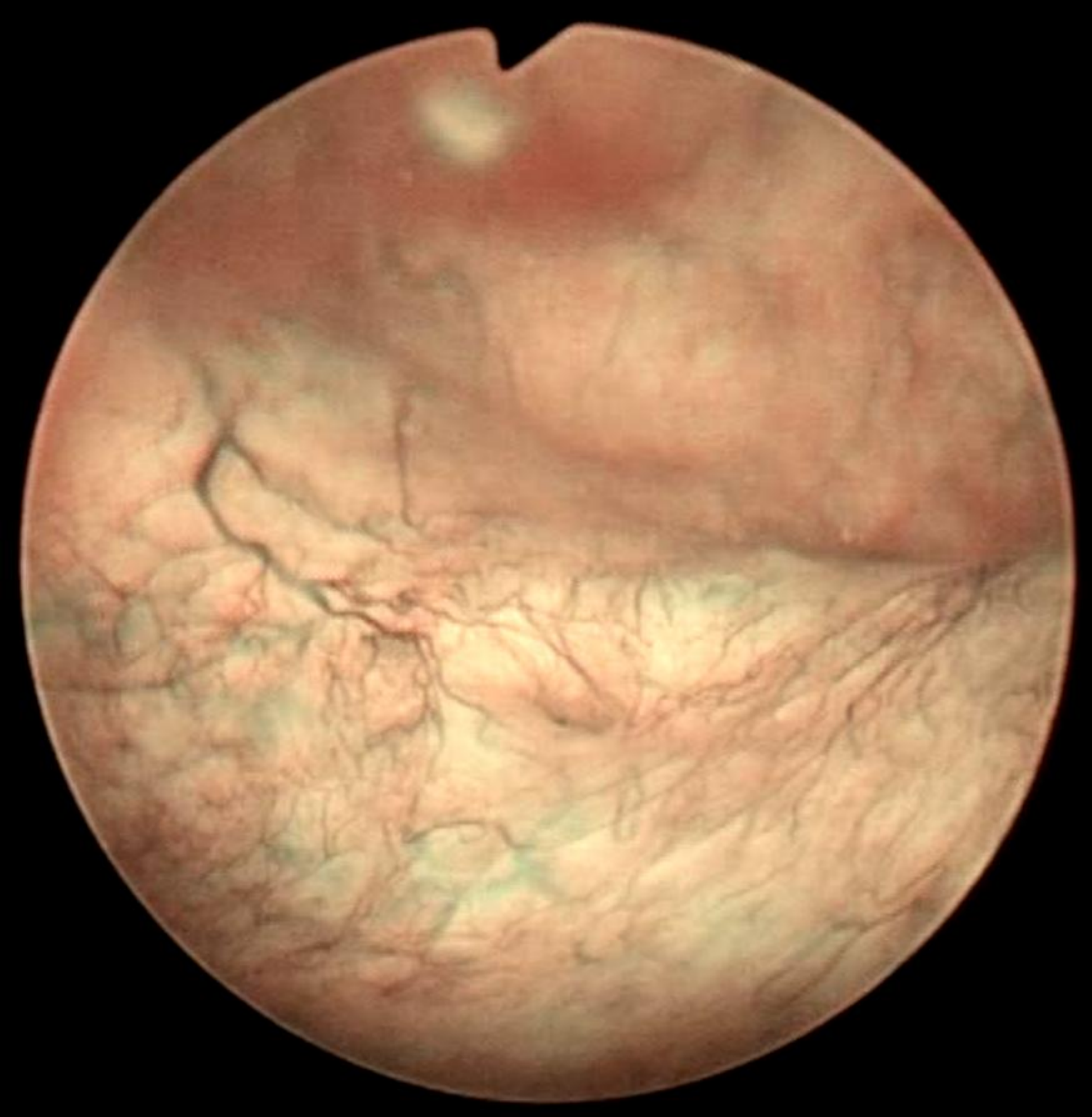}
        \includegraphics[width=\widthfig, height=\heighfig]{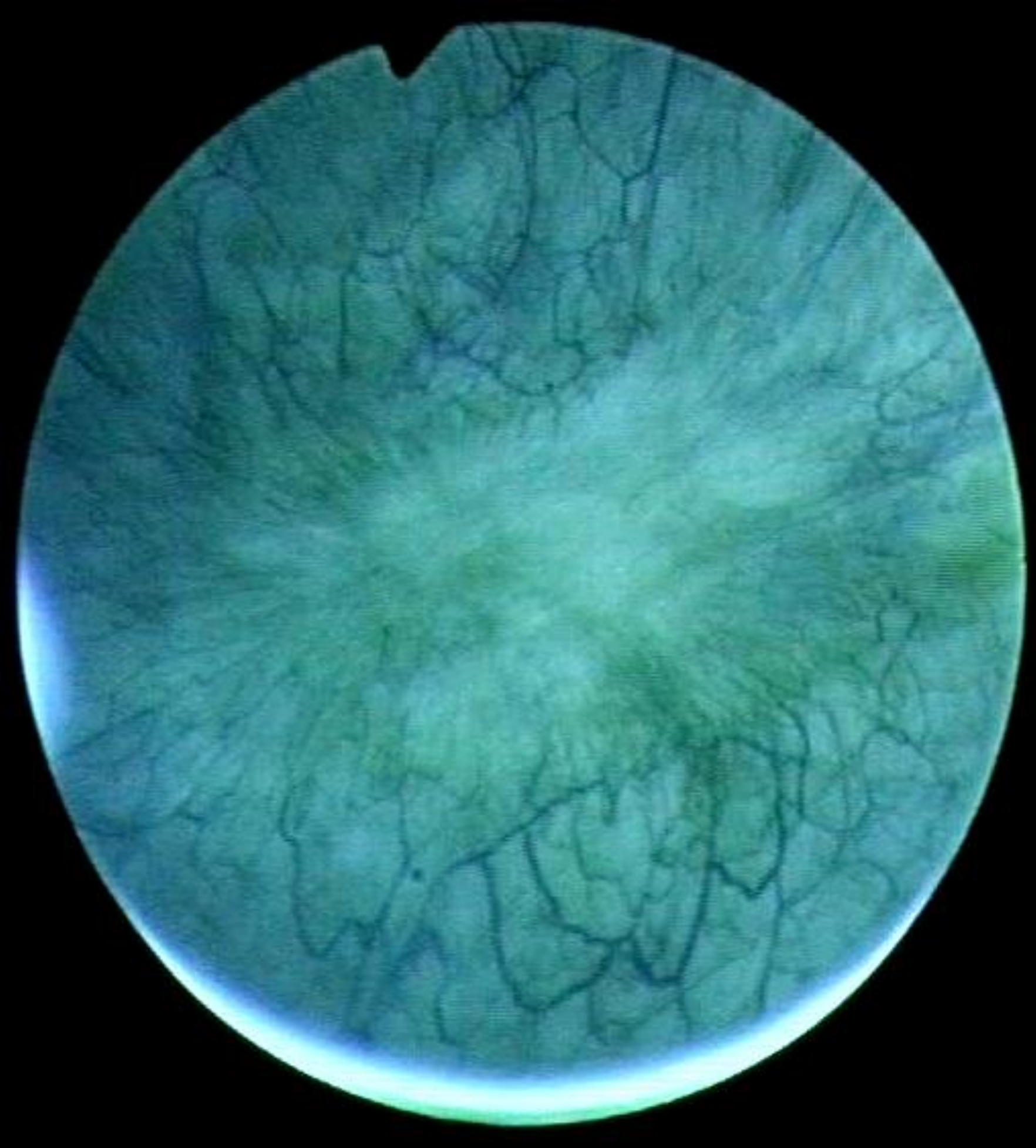}
         \caption{\footnotesize{}}
    \end{subfigure}
    \begin{subfigure}[b]{0.23\textwidth}
    \centering
        \includegraphics[width=\widthfig, height=\heighfig]{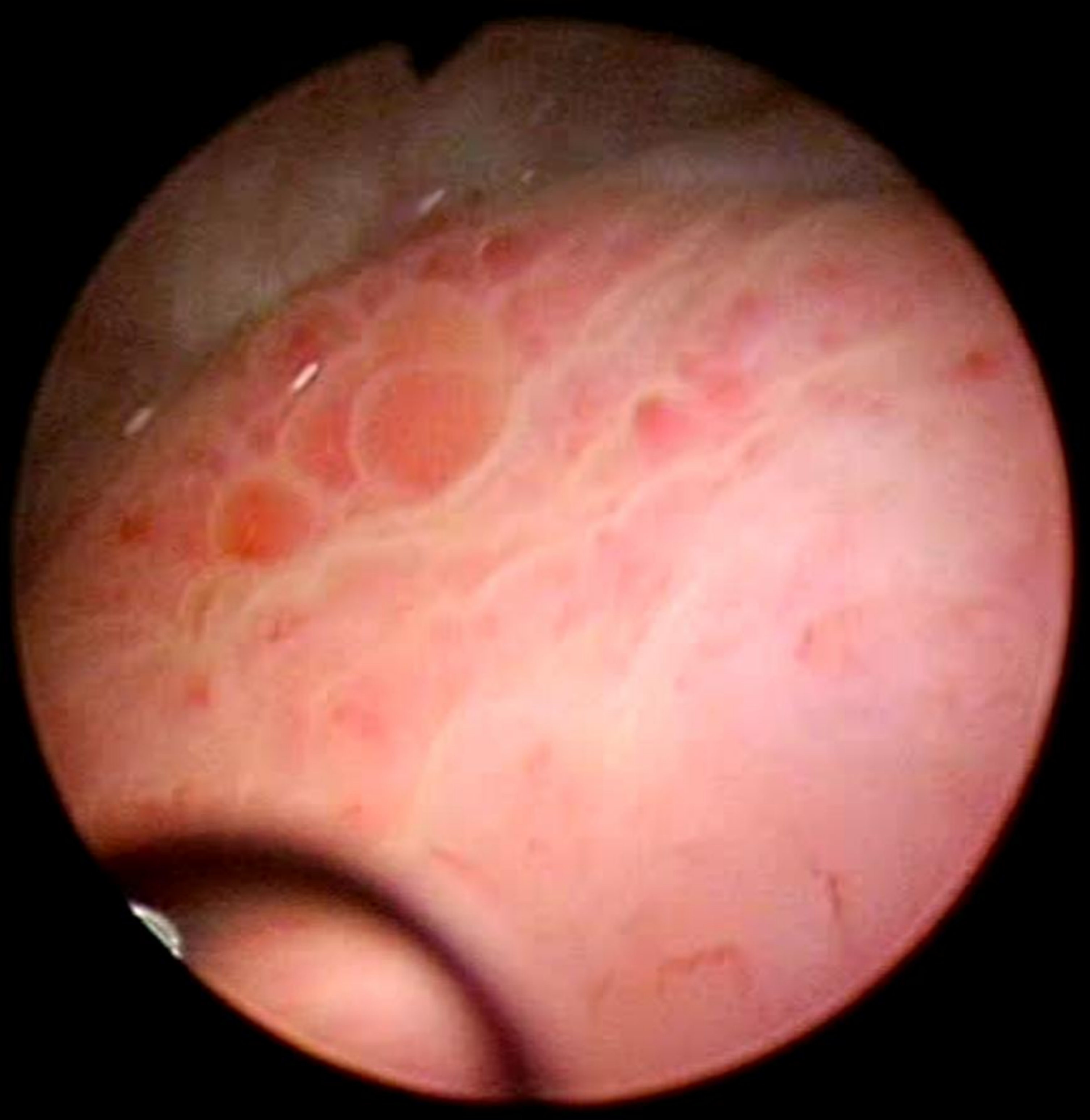} 
        \includegraphics[width=\widthfig, height=\heighfig]{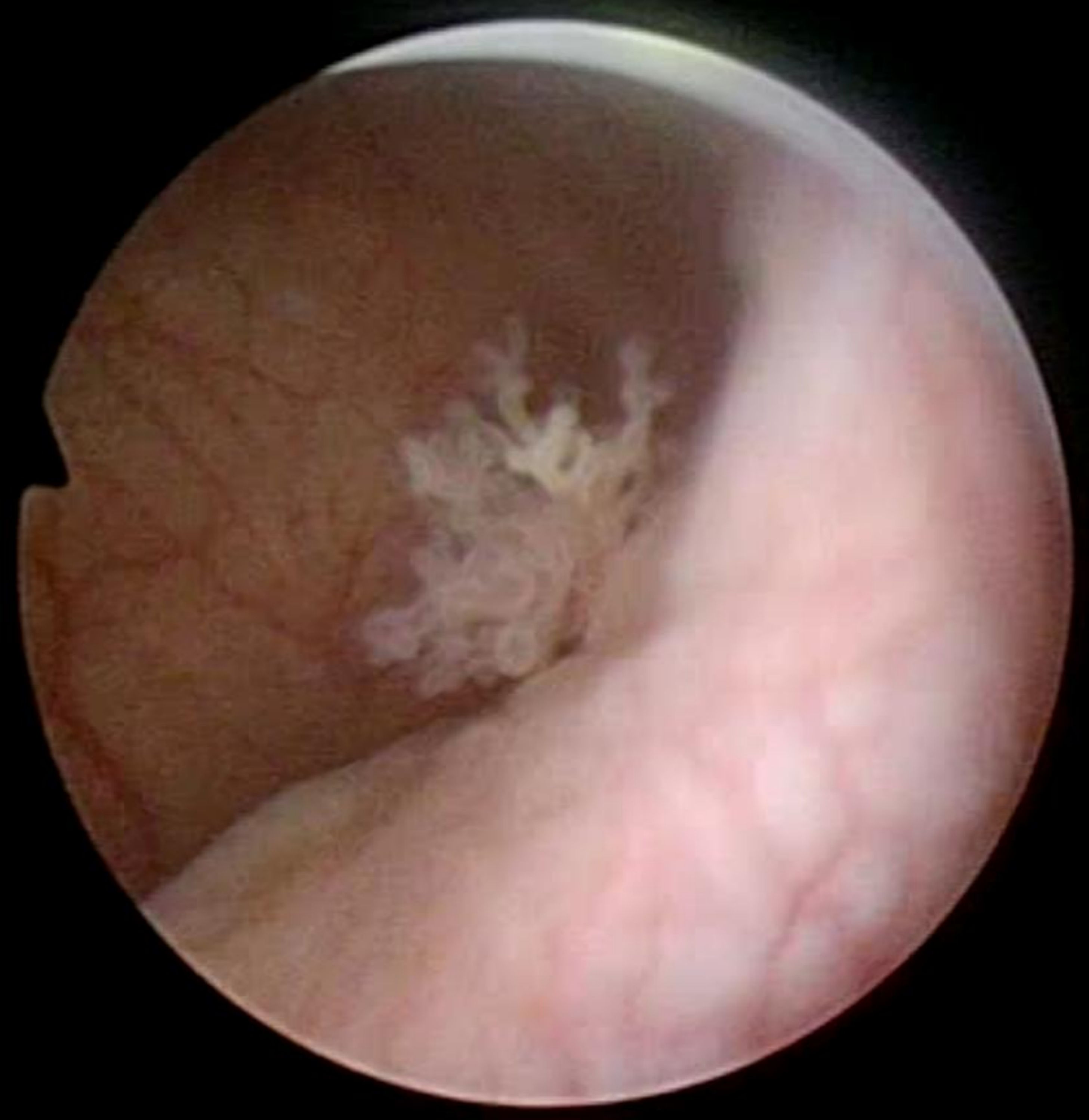} 
          
        \includegraphics[width=\widthfig, height=\heighfig]{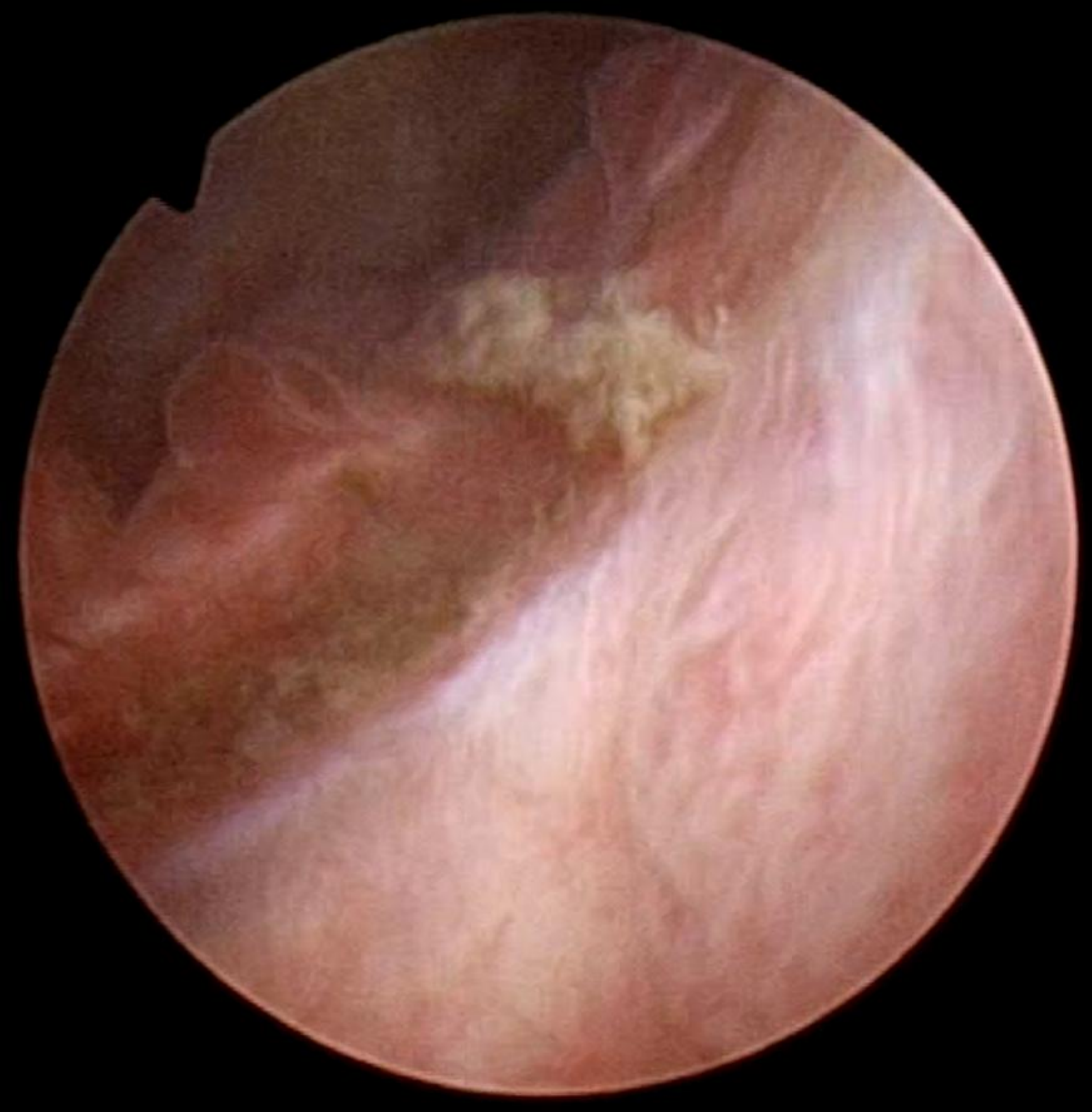}
        \includegraphics[width=\widthfig, height=\heighfig]{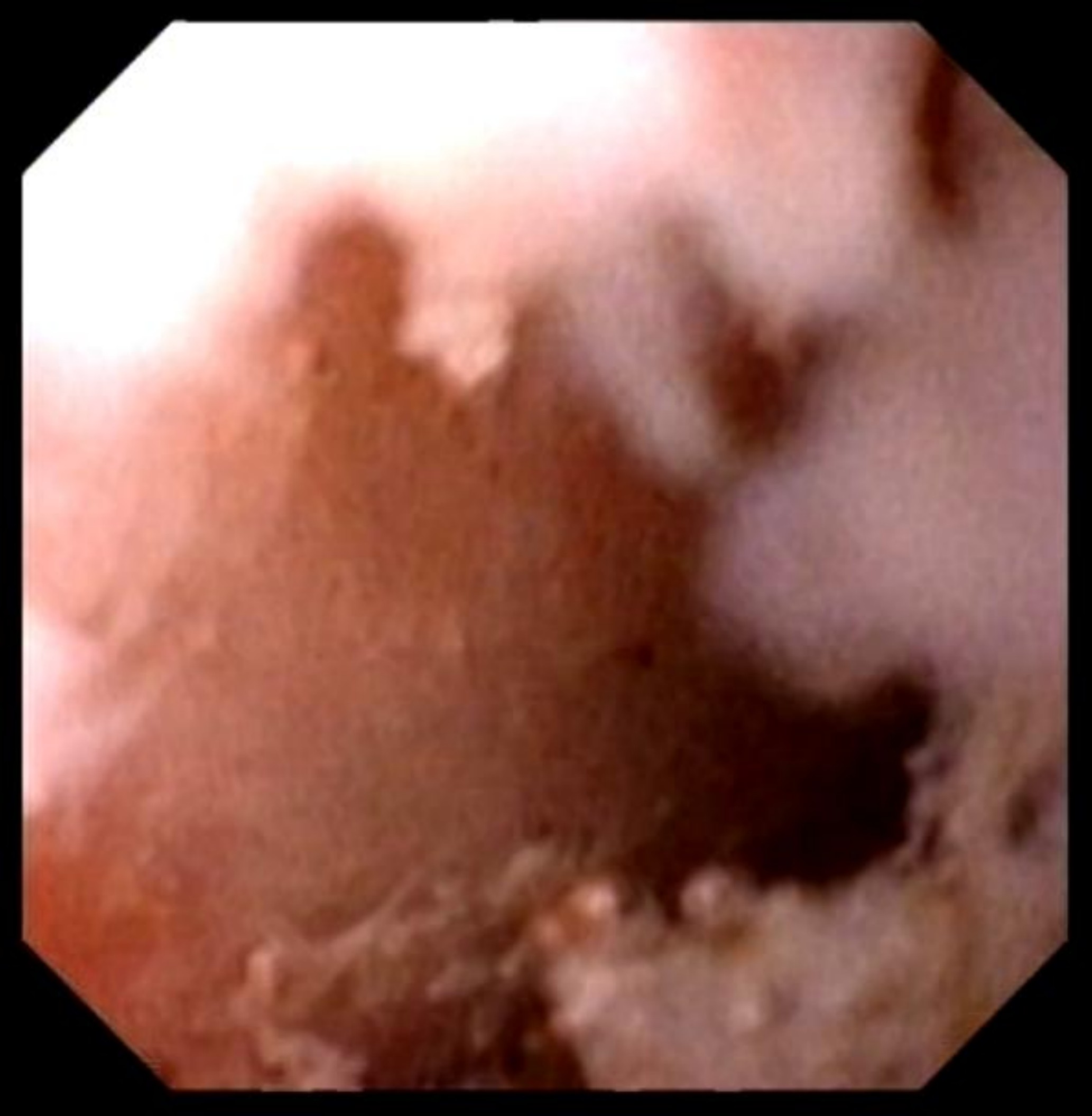}
    \caption{\footnotesize{}}
    \end{subfigure}
    \begin{subfigure}[b]{0.23\textwidth}
    \centering
        \includegraphics[width=\widthfig, height=\heighfig]{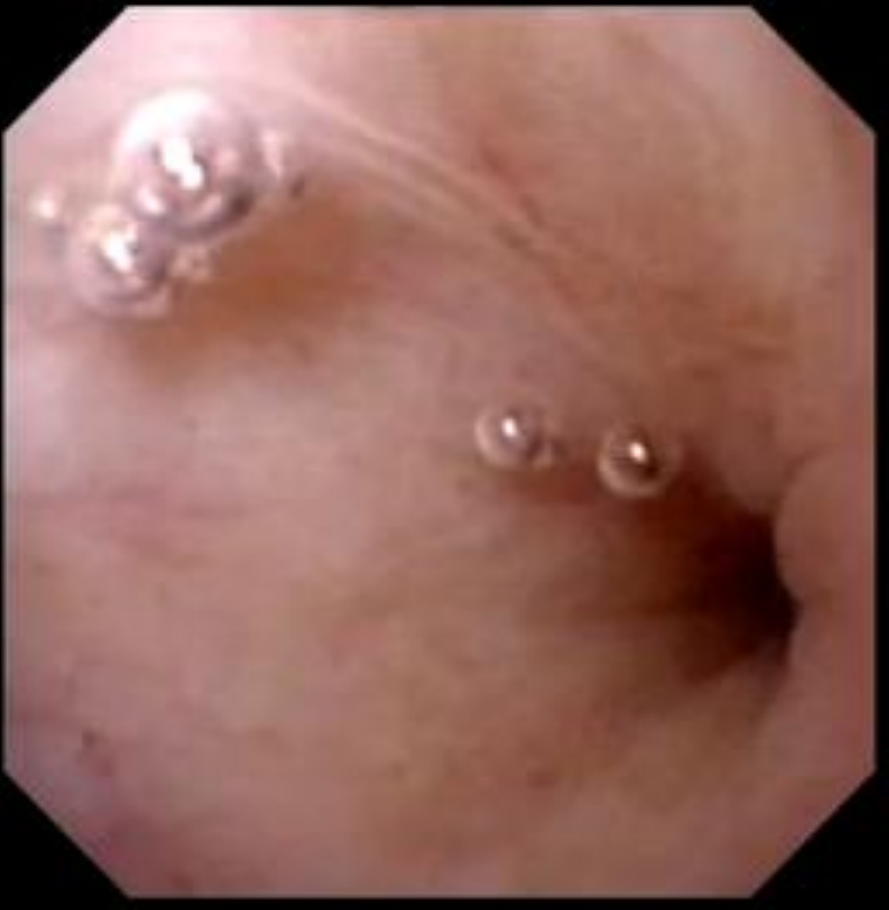} 
        \includegraphics[width=\widthfig, height=\heighfig]{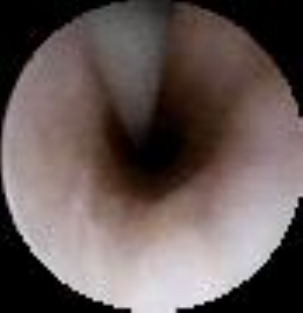} 
          
        \includegraphics[width=\widthfig, height=\heighfig]{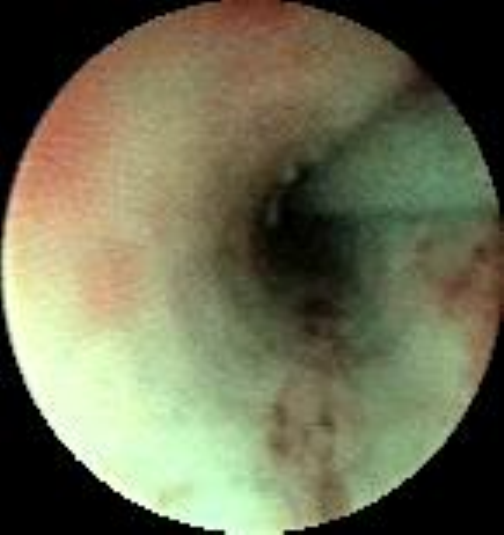}
        \includegraphics[width=\widthfig, height=\heighfig]{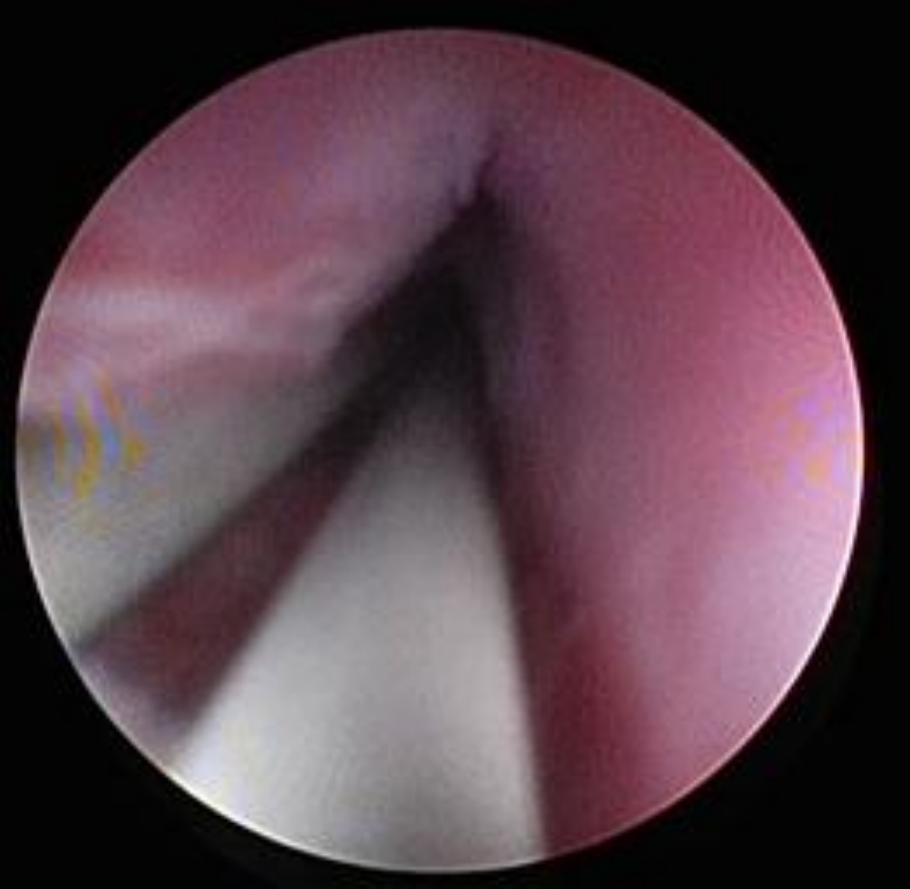}
         \caption{\footnotesize{}}
    \end{subfigure}
    \begin{subfigure}[b]{0.23\textwidth}
    \centering
        \includegraphics[width=\widthfig, height=\heighfig]{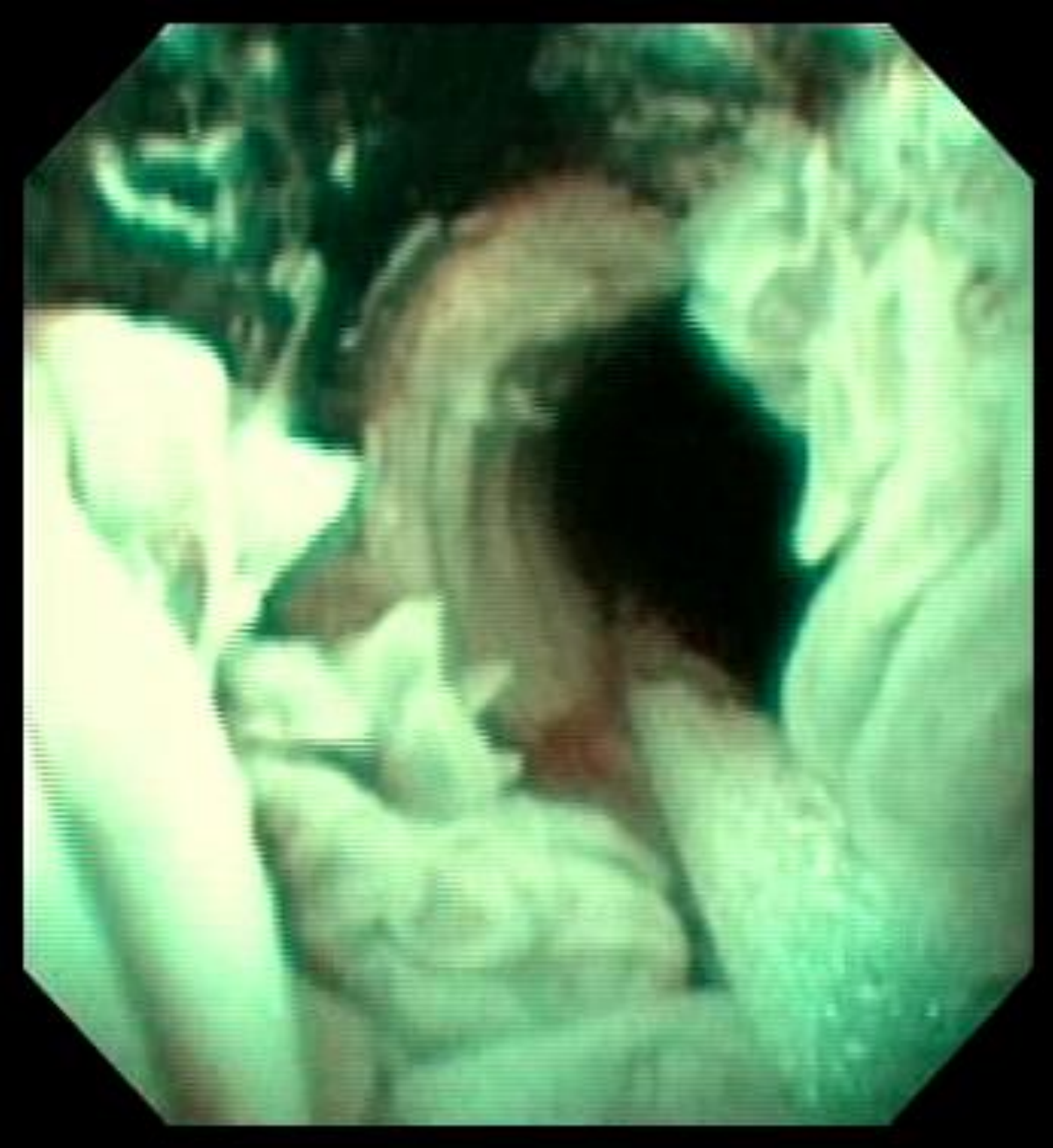} 
        \includegraphics[width=\widthfig, height=\heighfig]{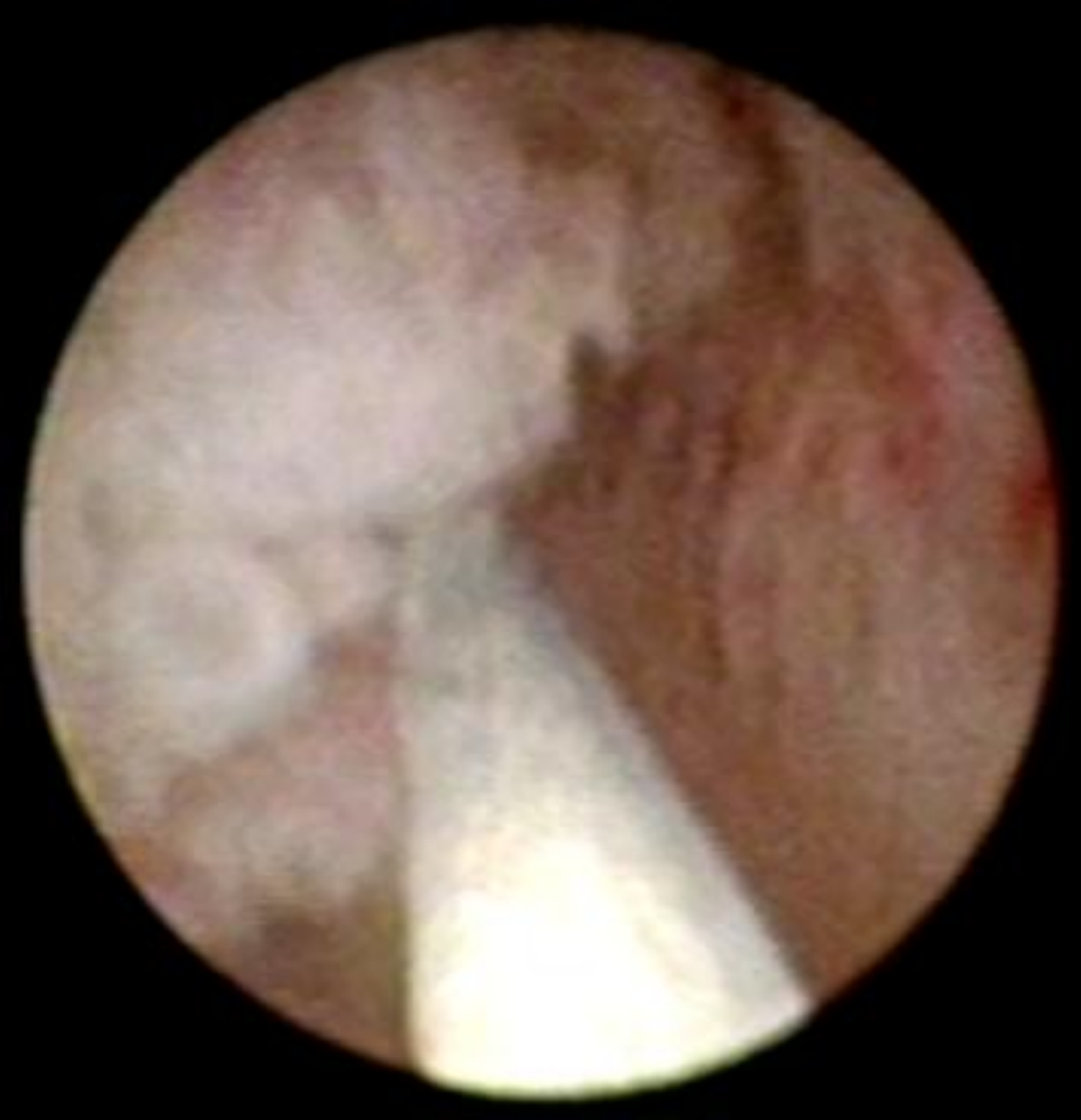} 
          
        \includegraphics[width=\widthfig, height=\heighfig]{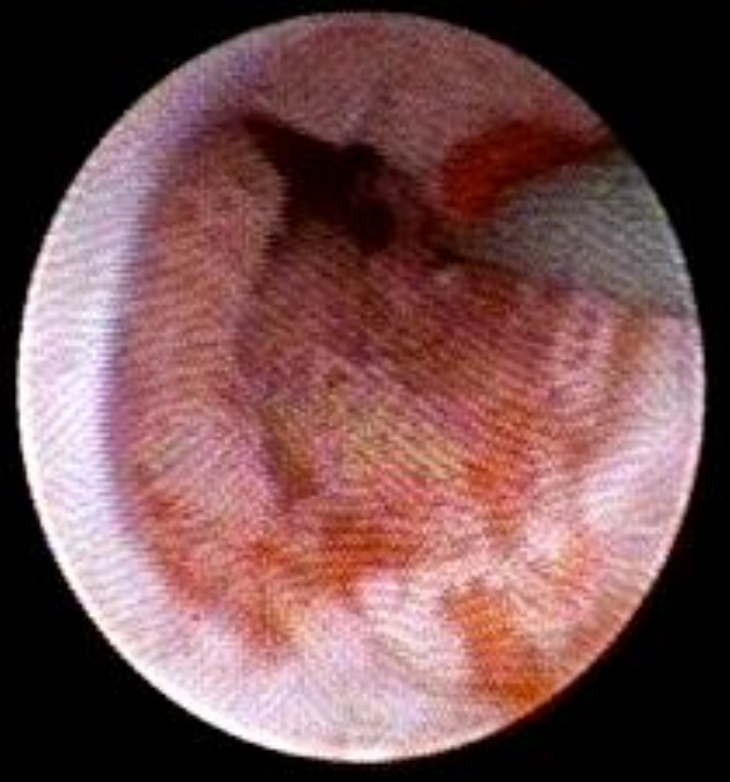}
        \includegraphics[width=\widthfig, height=\heighfig]{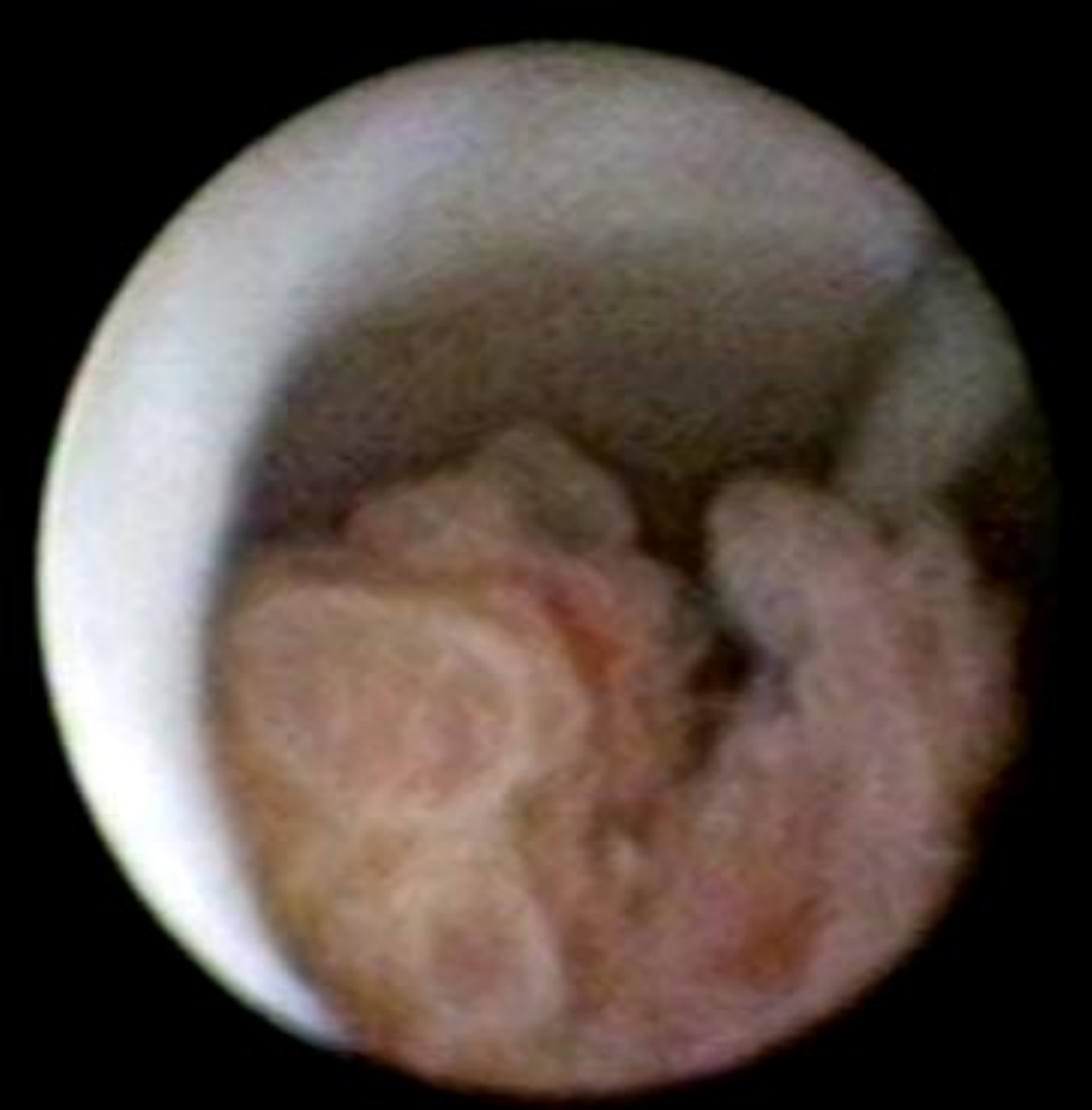}
    \caption{\footnotesize{}}
    \end{subfigure}

\caption{\footnotesize{Samples of endoscopic images in different organs in the urinary system. (a)-(b) images from the bladder. (c)-(d) images from the ureter and the kidney pelvis. (a), (c) correspond to samples of healthy tissue. (b), (d) samples of images with lesions.}}
\label{fig:sample_dataset}
\end{center}
\end{figure}

Few methods have been proposed for the detection and classification of tumors in the urinary system. 
This might be due to the lack of publicly available, annotated datasets.  
Shkolyar et al.~\cite{shkolyar2019augmented} propose a model based on CNNs for papillary and flat bladder tumor detection. 
Yang et al.~\cite{yang2020automatic} compared 3 different CNNs architectures as well as the DL platform EasyDL, training the models with WL cystoscopy images and showing that the use of DL-based methods can achieve accuracy values comparable to the ones of a surgical expert. Ikeda et al.~\cite{ikeda2020support} propose the use of transfer learning using GoogLeNet for the classification of endoscopic frames with and without nonmuscle-invasive BC. 
However, all these studies focus only on the bladder, missing the chance of validating their results in the whole urinary tract.  

As an initial step in the development of a fully automated CAD system to identify lesions in the urinary tract, we study the implementation of 3 CNNs. Considering the disparity in terms of occurrence of lesions in different organs in the urinary system, and the existing limitations in data collection and sharing, we study to which extent it is possible for a CNN to make inferences when it is trained on a similar domain, and if the use of transfer learning is suitable to achieve a better generalization in a different domain. 

We propose a two-steps transfer learning strategy where initially the CNNs are trained separately on ureteroscopy and cystoscopy data. 
After evaluating each network on its own domain and the other domain in which it was not originally trained (e.g. to identify lesions on cystoscopy images when the networks have been trained on ureteroscopy ones) we perform a second training stage. 
We change the training data to a different domain  i.e., the networks originally trained with cystoscopy are then trained with ureteroscopy images and the networks originally trained with ureteroscopy are trained with cystoscopy images. 
We compare this approach to the one of training the networks initially with both datasets. 

Determining to which extend the former approach is valid of relevance, since full annotated datasets are not always available. 

\section{PROPOSED METHODS}
\label{sec:methods}
\begin{figure}[tbp]
    \centering
    \includegraphics[width=0.45\textwidth]{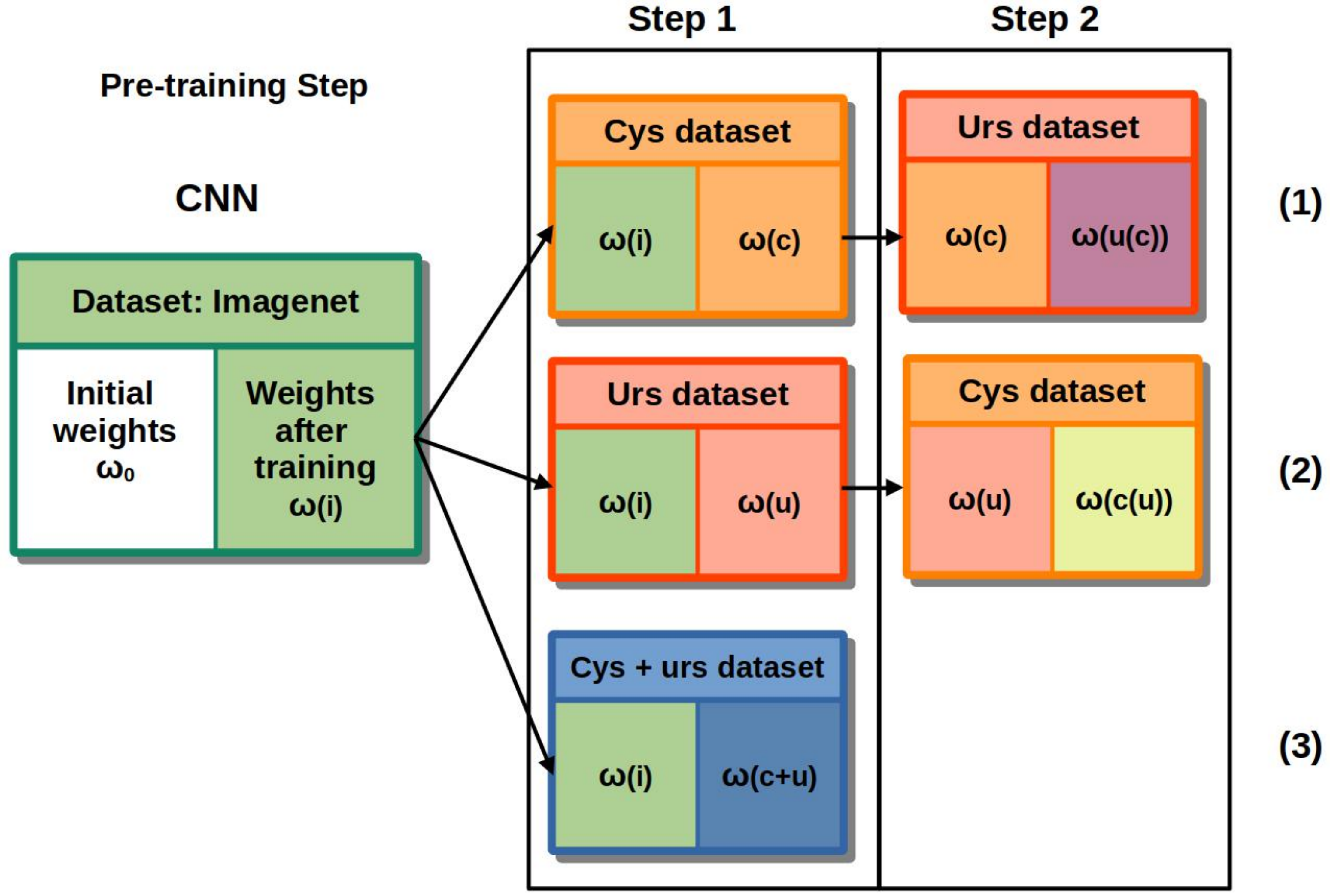}
    \caption{\footnotesize{Training strategy implemented for the 3 training cases. Each sub-table represents: 
    [starting weights (loaded ones) $|$ "output" weights (weights after training)]}}
    \label{fig:diagram_method}
\end{figure}

In this work, we investigated the use of transfer learning using different CNN architectures trained on data from bladder and upper urinary tract. 
Three state state-of-the-art architectures built under different paradigms for image classification, and whose weights after being trained on Imagenet dataset are publicly available, were considered. 

\begin{itemize}
    \item \textbf{VGG-16} is a 16-layer CNN model which has a feed-forward architecture consisting of 13 convolutional layers and 5 max-pooling layers~\cite{vgg_16_paper}. 
    The architecture starts with a convolutional layer with 64 kernels and this number is doubled after each pooling operation until it reaches 512 kernels. 
    The pooling layers are placed after selected convolutional layers in order to reduce dimension in the activation maps and hence  of the number of parameters that the CNN needs to learn.  
    
    \item \textbf{Inception V3} uses an architectural block called inception module, it consists of convolutional kernels with different sizes (1x1, 3x3 and 5x5) that are connected in parallel. 
    The use of different kernel sizes allows the identification of image features at different scales~\cite{InceptionV3}. 
    
    \item \textbf{ResNet50} was presented during the ILSCRC 2015 classification challenge obtaining the 1st place. The core idea of this model is the use of residual blocks to deal with the vanishing gradient problem, these are formed by 3 convolutional blocks in which a short skip connection is placed between the input and the output of each residual block~\cite{he2016deep}. 
    
\end{itemize}

On top of the last layer of the networks 3 Fully Connected (FC) layers, of size 2048, 1024 and 2, were added. For the first 2 FC layers, the activation function was $ReLu$ while in the case of the last layer $softmax$ was used.

\begin{figure*}[tbp]
    \centering
    \begin{subfigure}[b]{0.3\textwidth}
        \centering
        \includegraphics[width=0.9\textwidth]{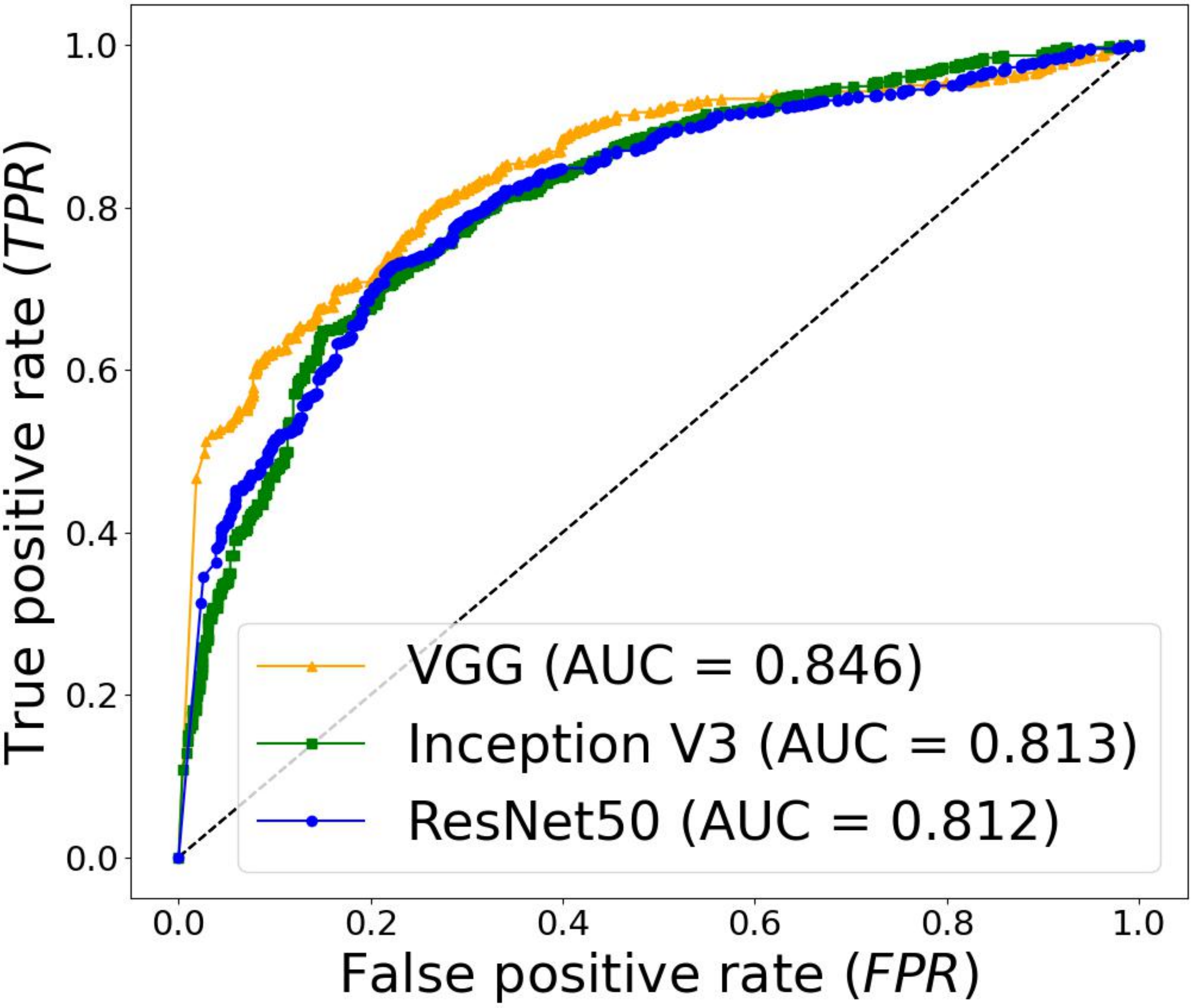}
        \caption{\footnotesize{Cystoscopy dataset}}
    \end{subfigure}
    \begin{subfigure}[b]{0.3\textwidth}
        \centering
        \includegraphics[width=0.9\textwidth]{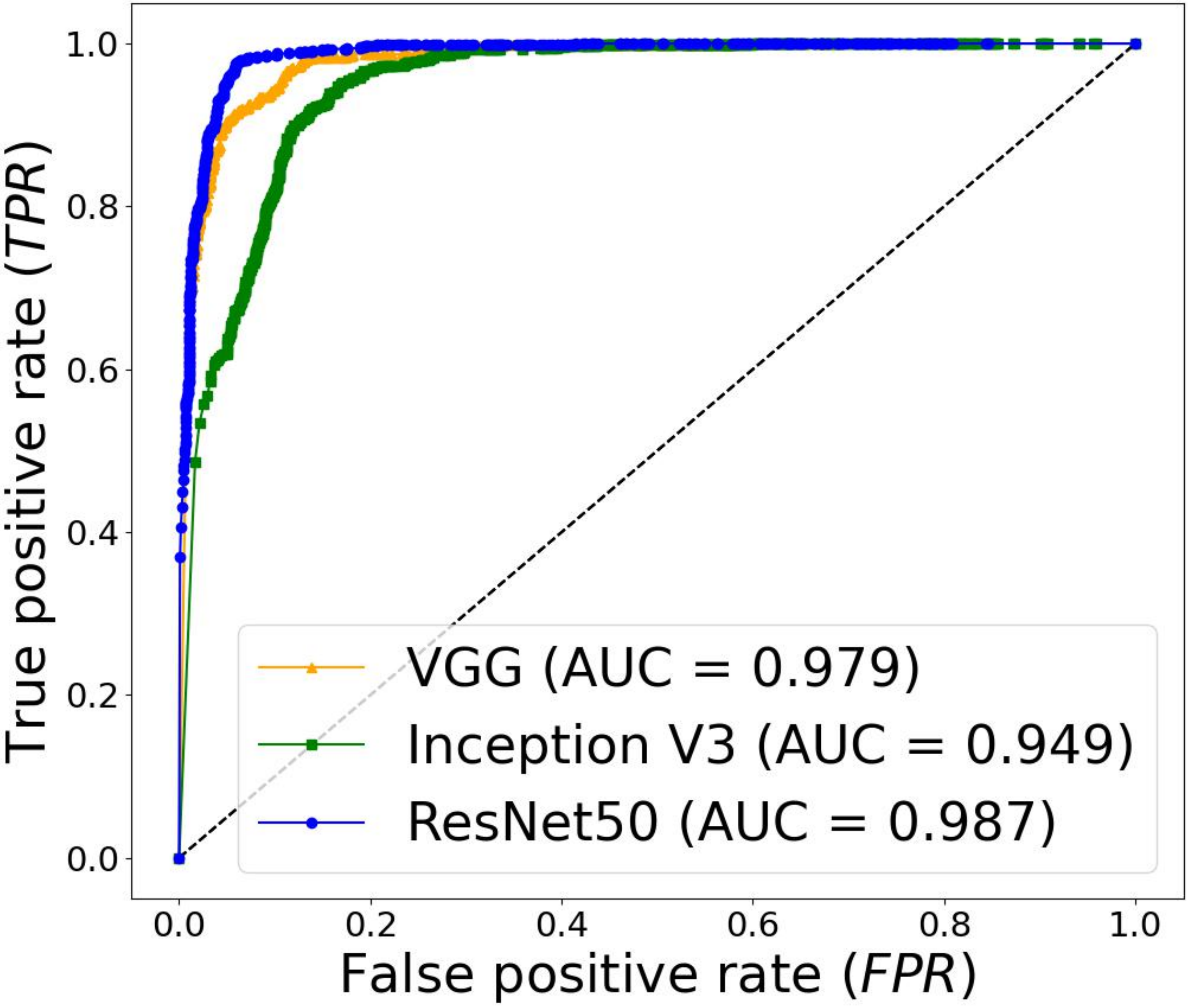}
        \caption{\footnotesize{Ureteroscopy dataset}}
    \end{subfigure}
    \begin{subfigure}[b]{0.3\textwidth}
        \centering
        \includegraphics[width=0.9\textwidth]{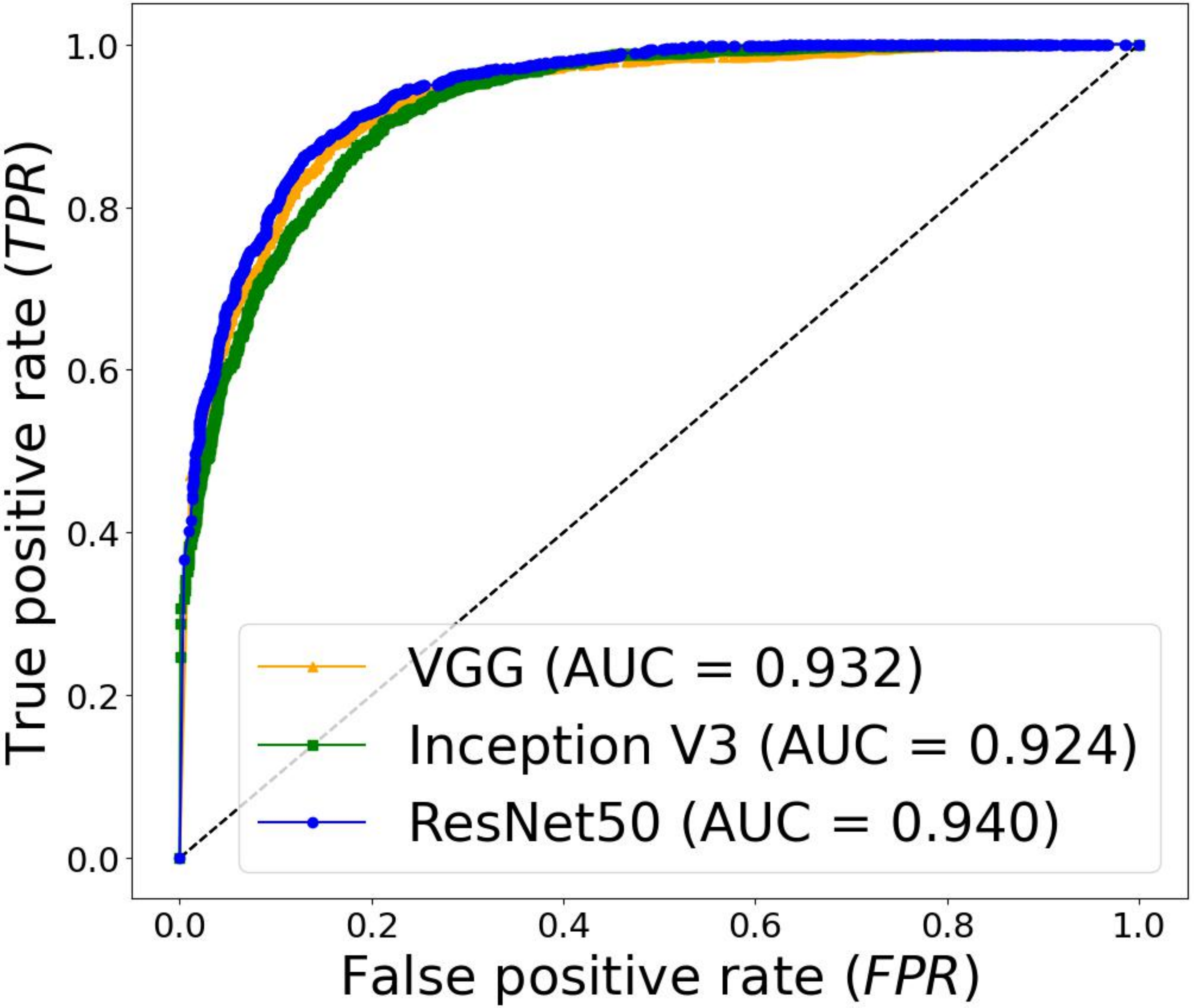}
        \caption{\footnotesize{Ureteroscopy + cystoscopy dataset}}
    \end{subfigure}
    \caption{\footnotesize{ROC curves obtained with the different models in the 3 datasets. The networks were trained and tested on data of the same domains.}}
    \label{fig:roc_curve_datasets}
\end{figure*}

\subsection{Dataset}
\label{sec:data}

\begin{table}[tbp]
    \centering
    \caption{\footnotesize{Dataset of the images collected. Two types of procedures were considered: ureteroscopy (urs) and cystoscopy (cys), and two light modalities; White Light Imaging (WLI) and Narrow Band Imaging (NBI).}}
    \begin{tabular}{c|c|c|c|c|c|c}
        & cases & \multicolumn{2}{c|}{lesion} & \multicolumn{2}{c|}{no lesion} & \\ \hline \hline
              &         & NBI & WLI  & NBI & WLI   & Total \\ \hline
        cys   &  11     & 337 & 906  & 298 & 1,346 & 2,887  \\
        urs   &  13     & 28  & 1801 & 227 & 1158  & 3,214 \\ \hline \hline
        Total &  24     & \multicolumn{2}{c|}{3,072}  & \multicolumn{2}{c|}{3,029} & 6,101
    \end{tabular}
    \label{tab:dataset}
\end{table}


For this study, we extracted frames from videos of 24 patients undergoing cystoscopy and/or ureteroscopy procedures using WLI and NBI. 
Of these videos, 13 correspond to patients undergoing ureteroscopy and 11 cystoscopy. 
The composition of the dataset is summarized in Table~\ref{tab:dataset}.
The images in the dataset corresponds to endoscopic images of different organs in the urinary tract including urethra, bladder, ureter and kidneys. 

The extracted images were labeled into two classes (with lesion, and without lesion) under the supervision  of an expert surgeon.
Some of the images include several artifacts such as blood, floating debris, specular reflections, bubbles, motion blur, as shown in Fig.~\ref{fig:sample_dataset}, which are common to appear in clinical practice and were included in the dataset to have a realistic approach.  
The videos were acquired from the European Institute of Oncology (IEO) at Milan, Italy following the ethical protocol approved by the IEO and in accordance with the Helsinky Declaration. 

\newcolumntype{C}[1]{>{\centering\arraybackslash}p{#1}}

\subsection{Training}

Transfer learning was implemented in a two-steps fashion. 
A diagram depicting the training strategy is shown in Fig.~\ref{fig:diagram_method}. The following 3 training scenarios were performed. 
\noindent
1)~During step 1, the CNN was trained only on cystoscopy data and tested on cystoscopy and ureteroscopy data. During step 2, only ureteroscopy images were used for training, and tested again in both cystoscopy and ureteroscopy images. 
\noindent
2)~The same procedure was carried out but inverting training data; during the first step ureteroscopy images were used for training and during the second stage only cystoscopy ones. 
\noindent
3)~Training was performed with cystoscopy and ureteroscopy images in one single step and the test was done using both types of images. 

For the 3 methods, at step 1 the initial weights of the networks were the ones obtained by previously training the CNN on the Imagenet dataset $\omega(i)$ as depicted in Fig.~\ref{fig:diagram_method}. At the second step, for case 1) the initial weights were the ones obtained by training the network in cystoscopy data $\omega(c)$; for case 2) the initial weights of the networks were the ones obtained by training the network in ureteroscopy images $\omega(u)$.

At both steps the CNNs were initially re-trained for 5 epochs with a learning rate of 0.0001. 
Subsequently, the weights from all the layers except the last 4 were frozen. These layers were trained for 30 epochs using Adam optimization algorithm with a learning rate of 0.001. 
A 3-fold cross validation strategy was used to estimate the performances of the models at each step. 




The models were implemented using \emph{Keras} with \emph{TensorFlow} as backend in Python 3.6 trained on a NVIDIA GeForce RTX 280 GPU.  

\subsection{Performance Metrics}

The performance evaluation of each of the tested CNNs was done by computing the True Positive Rate: $TPR = \frac{TP}{TP+FN}$, the False Positive Rate: $FPR = \frac{FP}{FP+TN}$ and the Area Under the Curve (AUC) of the Receiver Operating Characteristic (ROC) curve. 
Where $TP$ and $TN$ are the number of endoscopic images correctly classified as having lesions and no lesions, respectively, $FN$ and $FP$ are the amount of images missclassified.


\begin{figure}[tbp]
    \centering
    \includegraphics[width=0.45\textwidth]{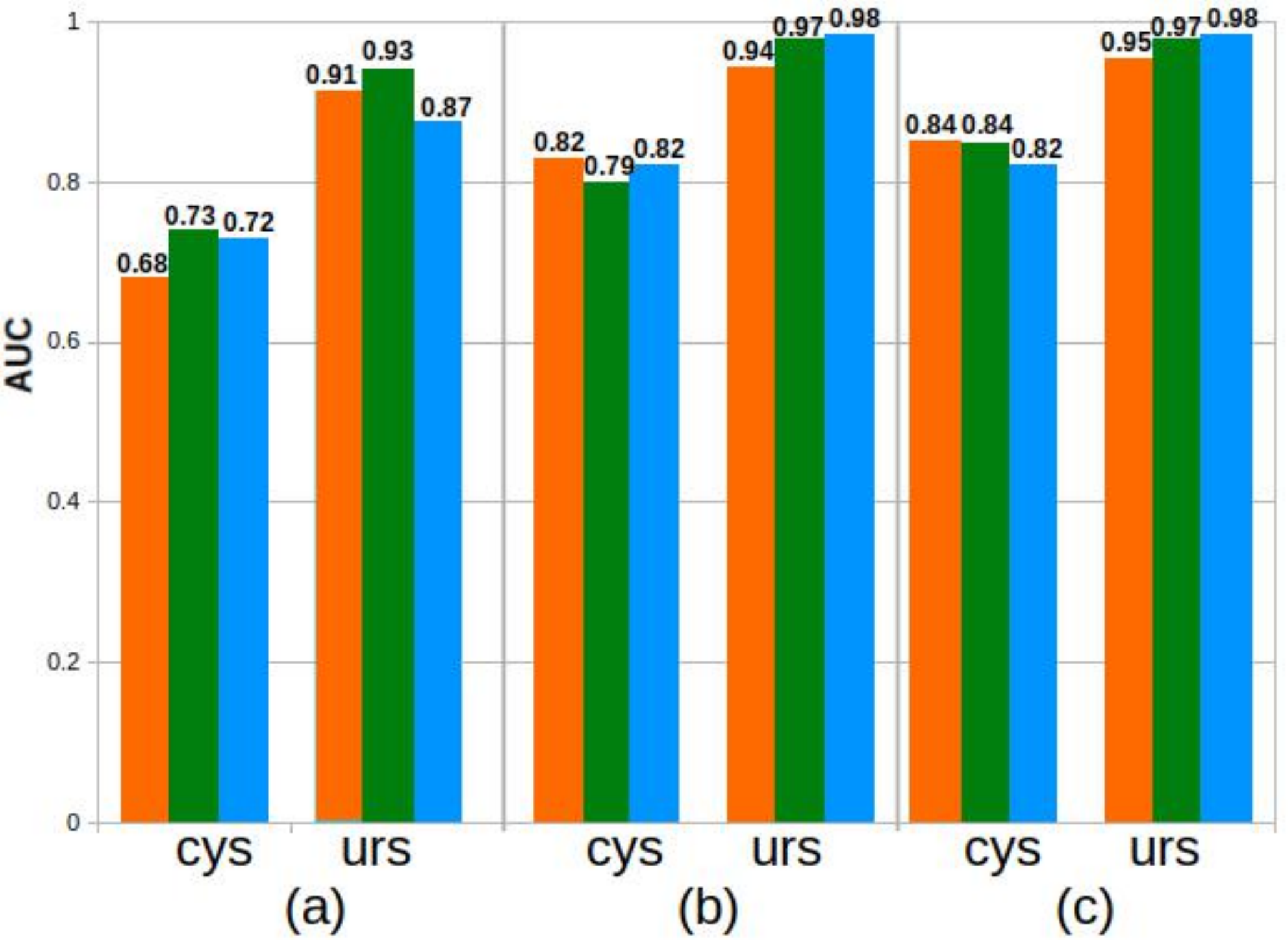}
    \caption{\footnotesize{Comparison of the {$AUC$} values obtained for each of the models trained. (orange: VGG16, green: Inception V3, blue: ResNet50). a)~Results of step 1~(see Fig.~\ref{fig:diagram_method}), when testing models on data of a different domain than the one for which they were trained on (i.e. predictions of cys data when trained on urs images, predictions of urs data when trained on cys).
    b)~Results of step 2, test data is of the same domain as the one of the second training. 
    c)~Separate results for the cystoscopy and ureteroscopy data on step 1 case (3). }}
    \label{fig:roc_curve_datasets_step12}
\end{figure}


\section{RESULTS}



The ROC curves obtained with each of the models on different datasets during the first step are shown in Fig.~\ref{fig:roc_curve_datasets}. 
In cystoscopy images the model that performed better is VGG with an AUC value of 0.846. In the case of ureteroscopy images, and the dataset composed by the combination of ureteroscopy and cystoscopy images, the model that obtains the better results is ResNet50 with AUC values of 0.987 and 0.938 respectively. 
Results comparing the performance before and after the second transfer learning step are shown Fig.~\ref{fig:roc_curve_datasets_step12}.

After completing Step 1 (see Fig.~\ref{fig:diagram_method}), Inception V3 is the model with best performances on data from a different domain than the one it was originally trained with. It obtains an AUC value of 0.895 when it is trained on cystoscopy images and tested on ureteroscopy ones, and 0.783 when trained on ureteroscopy and tested on cystoscopy images.



\begin{figure}
    \centering
    \begin{subfigure}[b]{0.14\textwidth}
    \centering
        \includegraphics[width=0.90\textwidth]{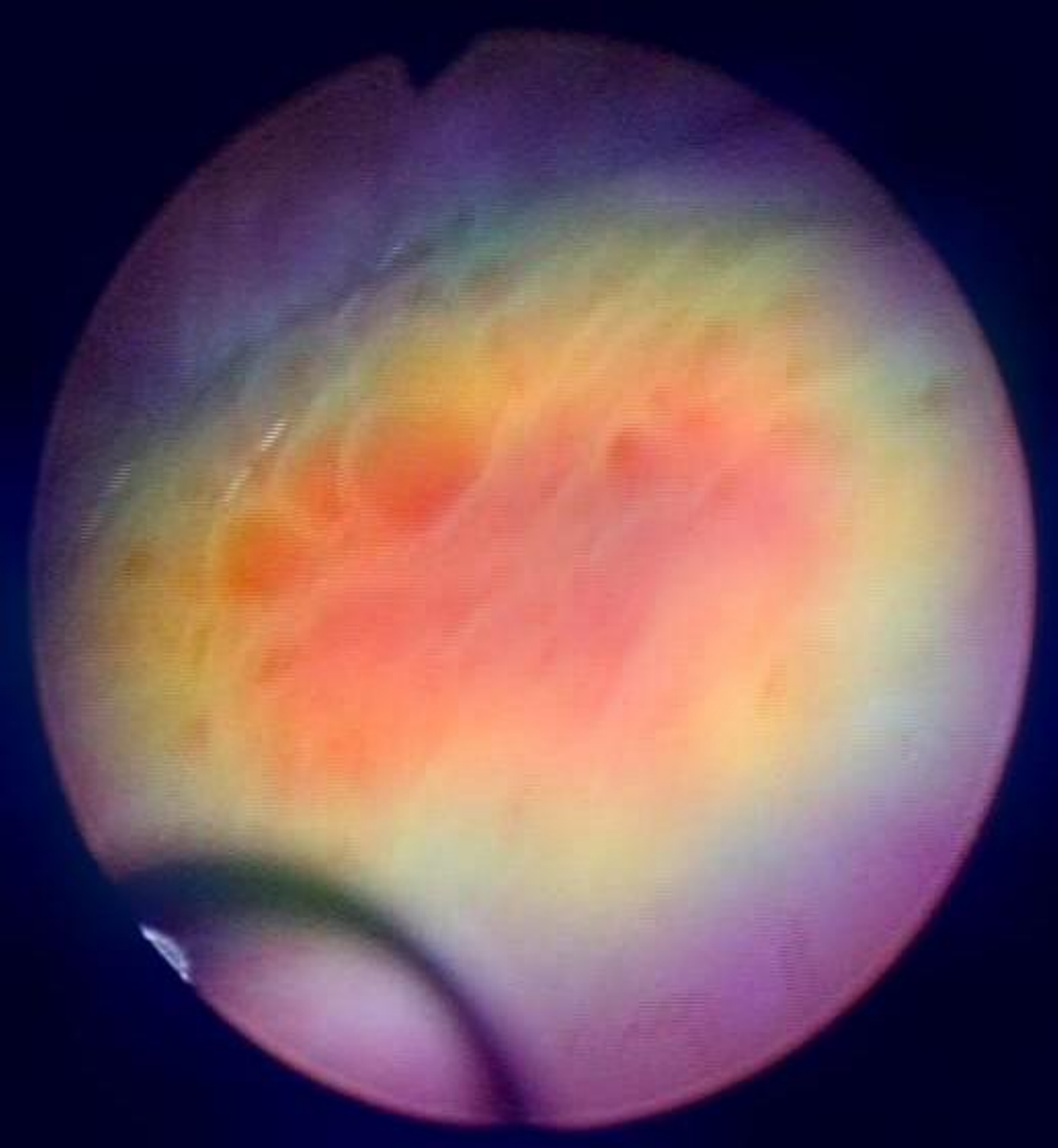} 
         \caption{\footnotesize{}}
    \end{subfigure}
    \begin{subfigure}[b]{0.14\textwidth}
    \centering
        \includegraphics[width=0.90\textwidth]{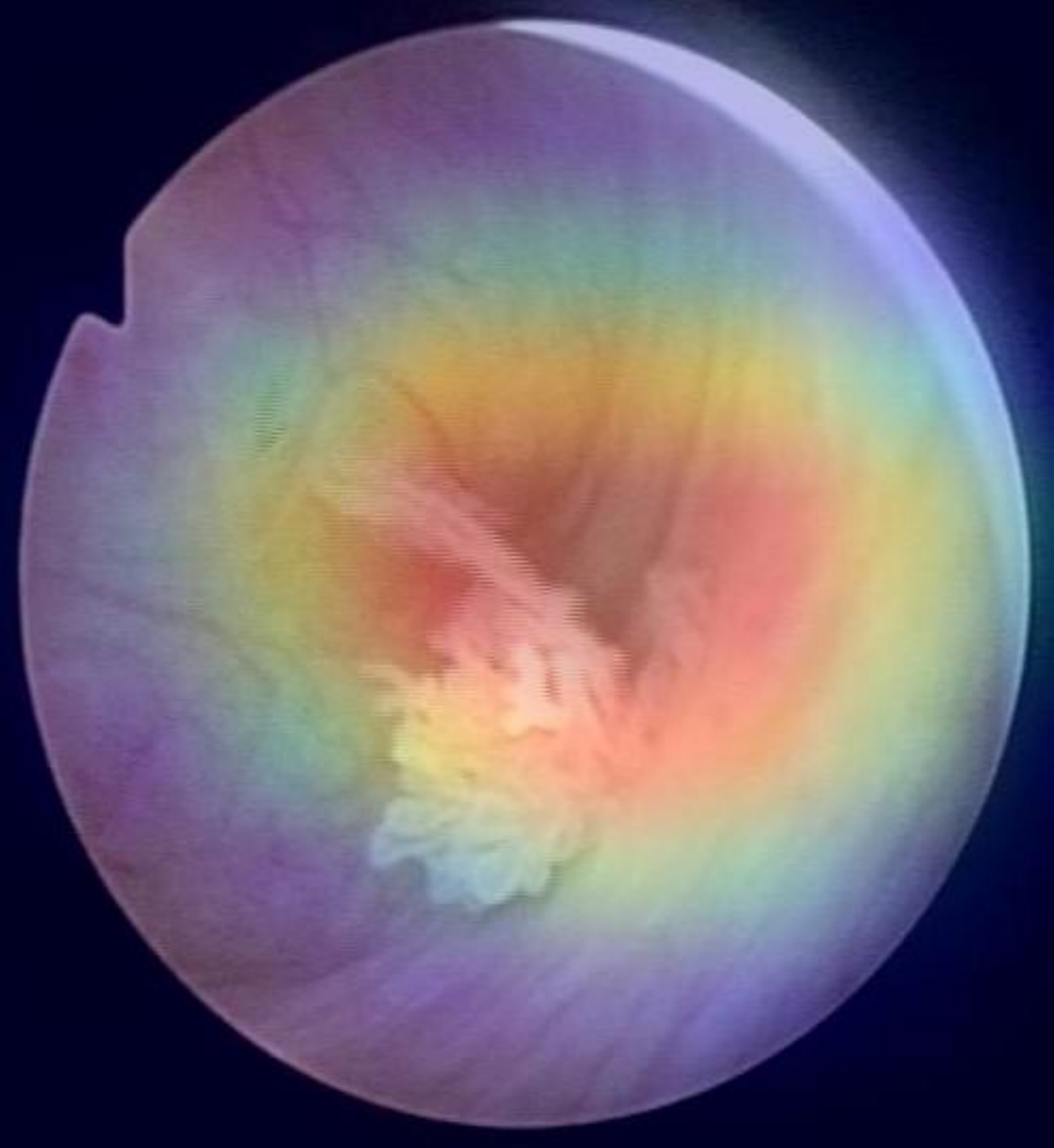} 
         \caption{\footnotesize{}}
    \end{subfigure}
    \begin{subfigure}[b]{0.14\textwidth}
    \centering
        \includegraphics[width=0.90\textwidth]{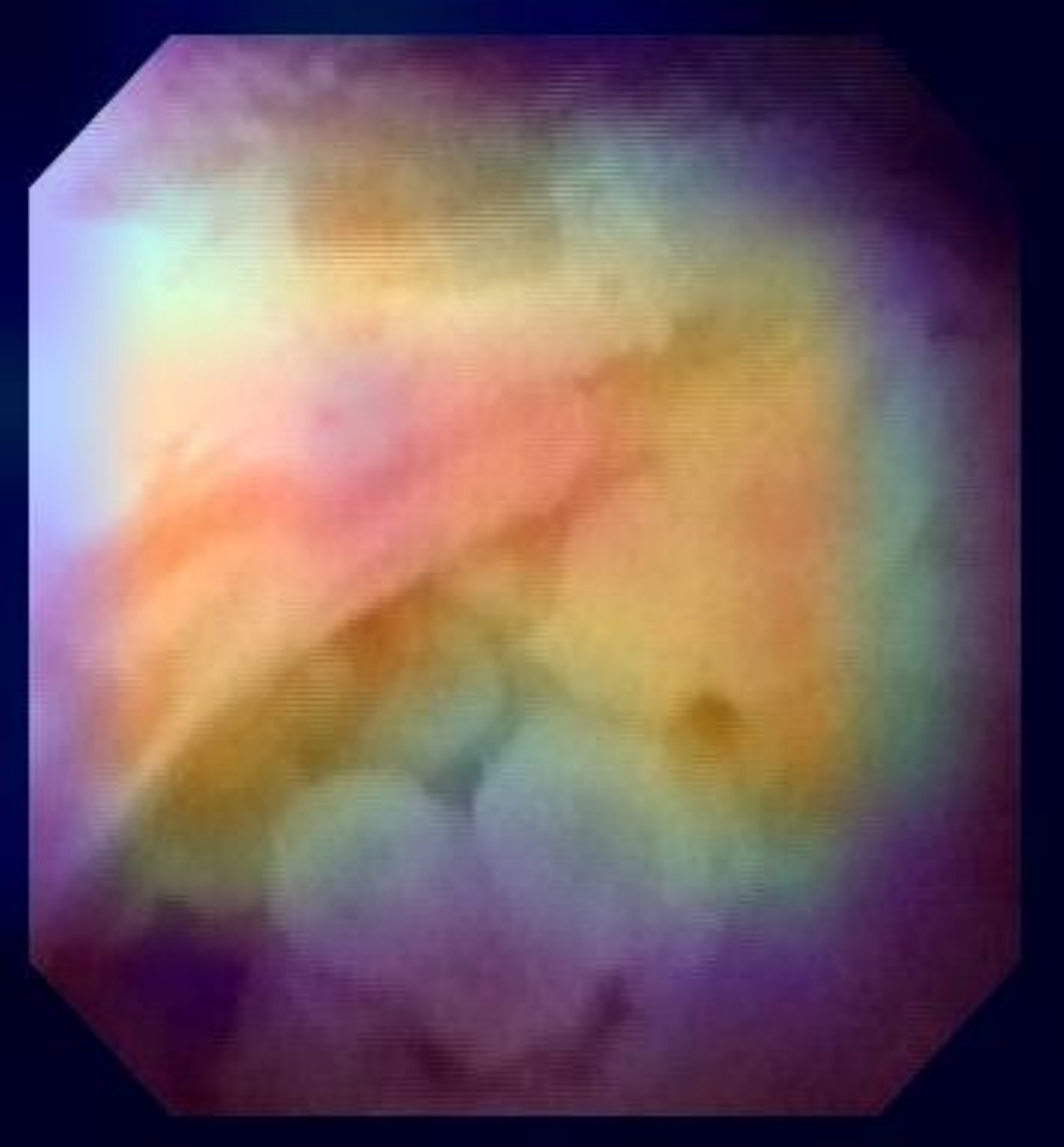} 
         \caption{\footnotesize{}}
    \end{subfigure}
    
    \begin{subfigure}[b]{0.14\textwidth}
    \centering
        \includegraphics[width=0.90\textwidth]{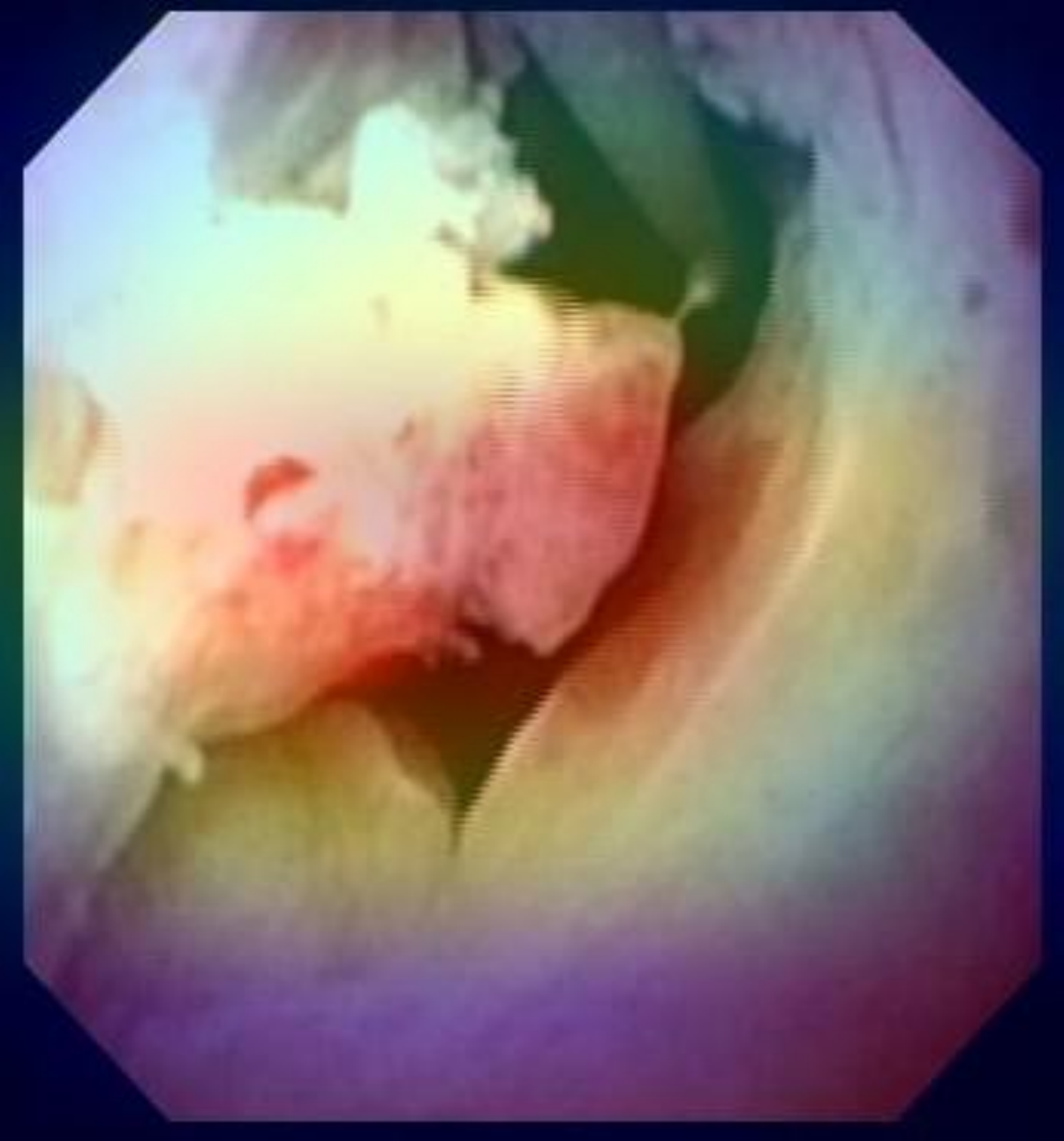} 
         \caption{\footnotesize{}}
    \end{subfigure}
    \begin{subfigure}[b]{0.14\textwidth}
    \centering
        \includegraphics[width=0.90\textwidth]{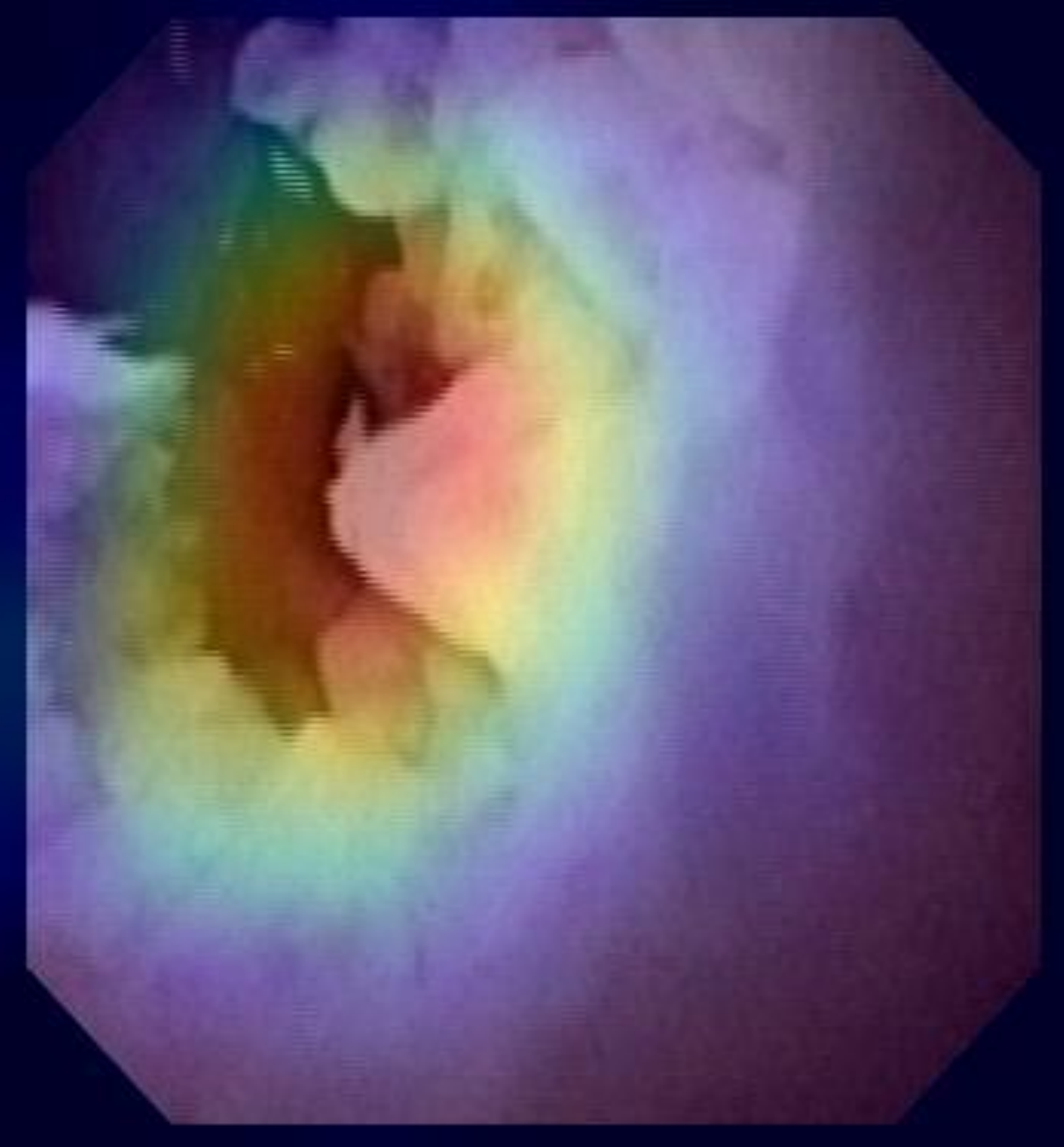} 
         \caption{\footnotesize{}}
    \end{subfigure}
    \begin{subfigure}[b]{0.14\textwidth}
    \centering
        \includegraphics[width=0.90\textwidth]{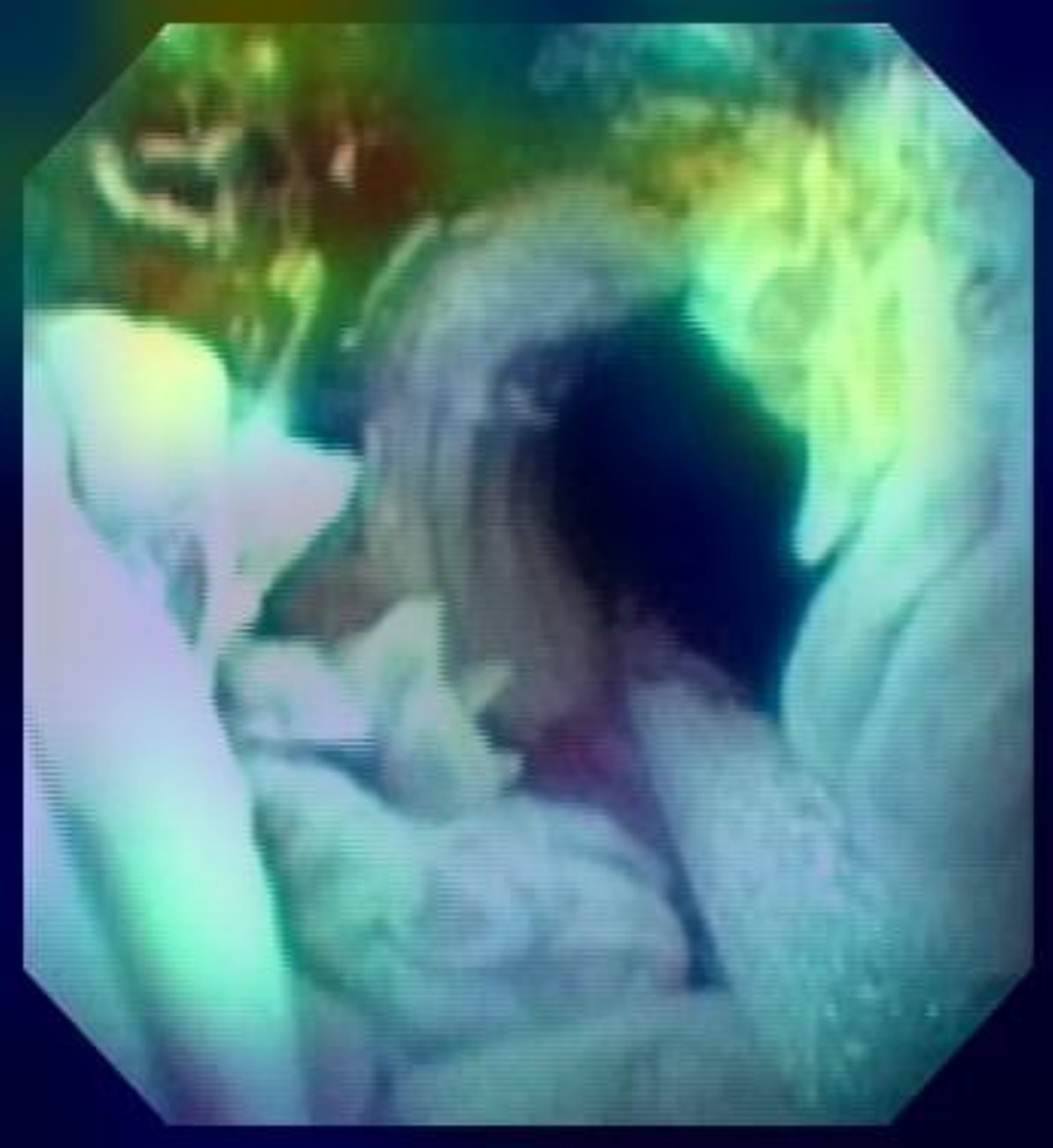} 
         \caption{\footnotesize{}}
    \end{subfigure}
    \caption{\footnotesize{Image samples where a
    class activation heat-map was added to show the regions of the images that the models recognized as lesions. (a)-(c) correspond to images from bladder whereas (d)-(c) correspond to images from the upper urinary tract. The images (a),(d) show samples of predictions using as training data cystoscopy images. (b),(e) show samples of predictions using as training data ureteroscopy images. (c),(f) show samples of predictions of a combination of both datasets for training.}}
    \label{fig:grad_cam_result_samples}
\end{figure}

The models that show a better improvement from the first to the second transfer learning step are VGG in cystoscopy data, changing from a AUC value of 0.691 to 0.834 and ResNet50 in ureteroscopy images improving from an initial AUC value of 0.897 to 0.979 after the second step. 
Some samples images of the predictions in the test dataset, with the heap-map obtained using Grad-CAM for class activation visualization~\cite{selvaraju2017grad} to show the regions of the image that the model recognized as part of the lesions, are shown in Fig.~\ref{fig:grad_cam_result_samples}.

\section{DISCUSSION $\&$  CONCLUSION}

All the networks performed better on detecting lesions on ureteroscopy than cystoscopy images. However, this might be related to the fact that the former dataset is bigger and has more patient cases than the later one. 

For the models tested, the performance was higher when the networks were trained on cystoscopy images and tested on ureteroscopy images than the opposite scenario. 
Considering that the dataset of ureteroscopy is larger than the ureteroscopy one, a possible explanation to this result might be related to the anatomical properties of the upper urinary tract and the bladder. While one is a tubular structure which in case of lesions might look like an obstruction, in the bladder the tissue has more visual diversity. 

The use of a training dataset which comprehends both domains resulted in better performances, as expected, but performing a second stage of transfer learning achieves comparable performances. 
In general ResNet50 is the network that achieves the better performances in both scenarios. 
Future work will include pathological confirmation of the type of tissue as ground truth to perform a more detailed classification of tissue which can differentiate between tumor and other non-tumor kind of lesions.





\footnotesize{
\section*{ACKNOWLEDGMENT}
This work was supported by the ATLAS project. This project has received funding from the European Union’s Horizon 2020 research and innovation programme under the Marie Skłodowska-Curie grant agreement No 813782.}


\bibliographystyle{IEEEtran}
\bibliography{biblio} 

\begin{thebibliography}{10}
\providecommand{\url}[1]{#1}
\csname url@samestyle\endcsname
\providecommand{\newblock}{\relax}
\providecommand{\bibinfo}[2]{#2}
\providecommand{\BIBentrySTDinterwordspacing}{\spaceskip=0pt\relax}
\providecommand{\BIBentryALTinterwordstretchfactor}{4}
\providecommand{\BIBentryALTinterwordspacing}{\spaceskip=\fontdimen2\font plus
\BIBentryALTinterwordstretchfactor\fontdimen3\font minus
  \fontdimen4\font\relax}
\providecommand{\BIBforeignlanguage}[2]{{%
\expandafter\ifx\csname l@#1\endcsname\relax
\typeout{** WARNING: IEEEtran.bst: No hyphenation pattern has been}%
\typeout{** loaded for the language `#1'. Using the pattern for}%
\typeout{** the default language instead.}%
\else
\language=\csname l@#1\endcsname
\fi
#2}}
\providecommand{\BIBdecl}{\relax}
\BIBdecl

\bibitem{siegel2020cancer}
R.~L. Siegel, K.~D. Miller, and A.~Jemal, ``Cancer statistics, 2020,''
  \emph{CA: A Cancer Journal for Clinicians}, vol.~70, no.~1, pp. 7--30, 2020.

\bibitem{kausch2010photodynamic}
I.~Kausch, M.~Sommerauer, F.~Montorsi, A.~Stenzl, D.~Jacqmin, P.~Jichlinski,
  D.~Jocham, A.~Ziegler, and R.~Vonthein, ``Photodynamic diagnosis in
  non--muscle-invasive bladder cancer: a systematic review and cumulative
  analysis of prospective studies,'' \emph{European Urology}, vol.~57, no.~4,
  pp. 595--606, 2010.

\bibitem{cosentino2013upper}
M.~Cosentino, J.~Palou, J.~M. Gaya, A.~Breda, O.~Rodriguez-Faba, and
  H.~Villavicencio-Mavrich, ``Upper urinary tract urothelial cell carcinoma:
  location as a predictive factor for concomitant bladder carcinoma,''
  \emph{World Journal of Urology}, vol.~31, no.~1, pp. 141--145, 2013.

\bibitem{wason2020ureteroscopy}
S.~E. Wason and S.~W. Leslie, ``Ureteroscopy,'' \emph{StatPearls}, 2020,
  (Accessed 29-11-2020), \url{https://pubmed.ncbi.nlm.nih.gov/32809391/}.

\bibitem{zheng2012narrow}
C.~Zheng, Y.~Lv, Q.~Zhong, R.~Wang, and Q.~Jiang, ``Narrow band imaging
  diagnosis of bladder cancer: systematic review and meta-analysis,'' \emph{BJU
  international}, vol. 110, no. 11b, pp. E680--E687, 2012.

\bibitem{sylvester2006predicting}
R.~J. Sylvester, A.~P. Van Der~Meijden, W.~Oosterlinck, J.~A. Witjes,
  C.~Bouffioux, L.~Denis, D.~W. Newling, and K.~Kurth, ``Predicting recurrence
  and progression in individual patients with stage {T}a {T}1 bladder cancer
  using eortc risk tables: a combined analysis of 2596 patients from seven
  eortc trials,'' \emph{European Urology}, vol.~49, no.~3, pp. 466--477, 2006.

\bibitem{chou2017comparative}
R.~Chou, S.~Selph, D.~I. Buckley, R.~Fu, J.~C. Griffin, S.~Grusing, and J.~L.
  Gore, ``Comparative effectiveness of fluorescent versus white light
  cystoscopy for initial diagnosis or surveillance of bladder cancer on
  clinical outcomes: systematic review and meta-analysis,'' \emph{The Journal
  of Urology}, vol. 197, no.~3, pp. 548--558, 2017.

\bibitem{poon2020ai}
C.~C. Poon, Y.~Jiang, R.~Zhang, W.~W. Lo, M.~S. Cheung, R.~Yu, Y.~Zheng, J.~C.
  Wong, Q.~Liu, S.~H. Wong \emph{et~al.}, ``A{I}-doscopist: a real-time
  deep-learning-based algorithm for localising polyps in colonoscopy videos
  with edge computing devices,'' \emph{NPJ Digital Medicine}, vol.~3, no.~1,
  pp. 1--8, 2020.

\bibitem{patrini2020transfer}
I.~Patrini, M.~Ruperti, S.~Moccia, L.~S. Mattos, E.~Frontoni, and E.~De~Momi,
  ``Transfer learning for informative-frame selection in laryngoscopic videos
  through learned features,'' \emph{Medical \& Biological Engineering \&
  Computing}, pp. 1--14, 2020.

\bibitem{shkolyar2019augmented}
E.~Shkolyar, X.~Jia, T.~C. Chang, D.~Trivedi, K.~E. Mach, M.~Q.-H. Meng,
  L.~Xing, and J.~C. Liao, ``Augmented bladder tumor detection using deep
  learning,'' \emph{European Urology}, vol.~76, no.~6, pp. 714--718, 2019.

\bibitem{yang2020automatic}
R.~Yang, Y.~Du, X.~Weng, Z.~Chen, S.~Wang, and X.~Liu, ``Automatic recognition
  of bladder tumours using deep learning technology and its clinical
  application,'' \emph{The International Journal of Medical Robotics and
  Computer Assisted Surgery}, p. e2194, 2020.

\bibitem{ikeda2020support}
A.~Ikeda, H.~Nosato, Y.~Kochi, T.~Kojima, K.~Kawai, H.~Sakanashi, M.~Murakawa,
  and H.~Nishiyama, ``Support system of cystoscopic diagnosis for bladder
  cancer based on artificial intelligence,'' \emph{Journal of Endourology},
  vol.~34, no.~3, pp. 352--358, 2020.

\bibitem{vgg_16_paper}
K.~Simonyan and A.~Zisserman, ``Very deep convolutional networks for
  large-scale image recognition,'' \emph{arXiv preprint arXiv:1409.1556}, 2014.

\bibitem{InceptionV3}
C.~Szegedy, V.~Vanhoucke, S.~Ioffe, J.~Shlens, and Z.~Wojna, ``Rethinking the
  inception architecture for computer vision,'' in \emph{IEEE Conference on
  Computer Vision and Pattern Recognition}, 2016, pp. 2818--2826.

\bibitem{he2016deep}
K.~He, X.~Zhang, S.~Ren, and J.~Sun, ``Deep residual learning for image
  recognition,'' in \emph{Proceedings of the IEEE Conference on Computer Vision
  and Pattern Recognition}, 2016, pp. 770--778.

\bibitem{selvaraju2017grad}
R.~R. Selvaraju, M.~Cogswell, A.~Das, R.~Vedantam, D.~Parikh, and D.~Batra,
  ``Grad-cam: Visual explanations from deep networks via gradient-based
  localization,'' in \emph{Proceedings of the IEEE international conference on
  computer vision}, 2017, pp. 618--626.

\end{thebibliography}

\addtolength{\textheight}{-12cm}   


\end{document}